\documentclass{article}
\usepackage{square_preprint,times}
\squarefinalcopy
\usepackage{bbm}
\usepackage[utf8]{inputenc} % allow utf-8 input
\usepackage[T1]{fontenc}    % use 8-bit T1 fonts
\usepackage{hyperref}       % hyperlinks
\usepackage{url}            % simple URL typesetting
\usepackage{booktabs}       % professional-quality tables
\usepackage{amsfonts}       % blackboard math symbols
\usepackage{nicefrac}       % compact symbols for 1/2, etc.

\usepackage{xcolor}         % colors
\usepackage{wrapfig}
\usepackage{subcaption}
\usepackage{graphicx}
\usepackage{amsthm}
\usepackage{amsmath}
\usepackage{mathtools}
\usepackage{subcaption}
\usepackage{xcolor}
\usepackage{apxproof}
\definecolor{lightgray}{cmyk}{0,0,0,.05}

\newlength\GreyboxOuterVspace
\newlength\GreyboxPadding
\newlength\GreyboxRule
\newcommand{\GreyboxFrameColor}{black}

\newcommand{\greybox}[1]{%
  \par\addvspace{\GreyboxOuterVspace}%
  \noindent\begingroup
  \setlength{\fboxsep}{\GreyboxPadding}%
  \setlength{\fboxrule}{\GreyboxRule}%
  \fcolorbox{\GreyboxFrameColor}{lightgray}{%
    \parbox{\dimexpr\linewidth-2\fboxsep-2\fboxrule\relax}{#1}%
  }
  \endgroup%
  \par\addvspace{\GreyboxOuterVspace}%
}

\usepackage{xspace}
\newtheorem{theorem}{Theorem}[section]

\newtheorem{remark}[theorem]{Remark}
\newcommand{\BfPara}[1]{{\noindent\bf#1.}\xspace}
\usepackage{bbding}
\usepackage{mdframed}
\usepackage{amsmath}

\usepackage{booktabs}
\usepackage{algorithm}
\usepackage{algorithmicx}
\usepackage{algpseudocode}
\usepackage{algpseudocode}
\usepackage{algpseudocode}
\usepackage{algcompatible}
\usepackage{multirow}
\usepackage{multicol}
\usepackage[export]{adjustbox}
\usepackage{array}
\usepackage{makecell}
\usepackage{amsthm}
\usepackage{tabularray}
\usepackage{braket}
\usepackage{amsmath}
\usepackage{color}
\usepackage{array}
\usepackage{amsthm}
\usepackage{float}
\usepackage{soul}
\usepackage{nccmath}
\usepackage{graphicx}
\usepackage{caption}
\usepackage{booktabs}
\usepackage{subcaption}
\usepackage{mwe}
\usepackage{amsmath,amsfonts}
\usepackage{textcomp}
\usepackage{xcolor}
\usepackage[utf8]{inputenc} % use UTF8 encoding

\usepackage{algpseudocode}
\usepackage{tcolorbox}
\usepackage{multirow}
\usepackage{mathtools}
\usepackage{graphicx}
\usepackage{algorithm}
\usepackage{algpseudocode}
\usepackage{amsmath,amssymb}
\usepackage{float}
\usepackage{tikz}
\usepackage{amssymb}
\usepackage{wrapfig}
\usepackage{pifont}

\AtBeginDocument{%
  }
\usepackage{enumitem}

\usepackage{listings}
\usepackage{xcolor}

\lstdefinestyle{pythonstyle}{
    language=Python,
    basicstyle=\ttfamily\scriptsize,
    keywordstyle=\color{blue},
    commentstyle=\color{gray},
    stringstyle=\color{teal},
    showstringspaces=false,
    breaklines=true,
    breakatwhitespace=true,
    frame=single,
    rulecolor=\color{black!30},
    numbers=left,
    numberstyle=\tiny\color{gray},
    tabsize=4,
    captionpos=b
}

\usepackage[normalem]{ulem}

\title{SQUARE: Structured Quantum \\ Representation Adapters as Compact \\ Quadratic Feature Maps for \\ Frozen Language Models}
\author{\bfseries Emily Jimin Roh$^{1}$, Hyojun Ahn$^{1}$, Hoyeong Lee$^{1}$,  Soohyun Park$^{2}$, \\
\textbf{Sung Whan Yoon$^{3}$, Vaneet Aggarwal$^{4}$, Joongheon Kim$^{1,5}$\thanks{Corresponding author: \texttt{joongheon@korea.ac.kr}}} \\
{\normalfont\small $^{1}$Korea University \quad $^{2}$Sookmyung Women's University \quad $^{3}$Ulsan National Institute of Science and Technology} \\
{\normalfont\small $^{4}$Purdue University \quad $^{5}$Seoul National University Hospital}}

\begin{document}

\maketitle
\lhead{Preprint}
\hypersetup{hidelinks}
\begin{abstract}
Frozen language models (LMs) are increasingly used as fixed feature extractors for downstream reranking, scoring, and preference modeling, raising a practical question: how should a compact module represent interactions among features in a fixed low-dimensional bottleneck? Common linear and low-rank adapters remain linear at the adaptation module itself, whereas explicit second-order alternatives introduce pairwise interactions through direct parameterization or predefined factorizations. We propose \textbf{SQUARE}, a \textbf{S}tructured \textbf{QUA}ntum \textbf{RE}presentation adapter that amplitude-encodes the bottleneck vector, applies a parameterized quantum circuit, and measures the resulting state. We show that each basis-probability feature is exactly a normalized quadratic form in the bottleneck coordinates, while the additional Pauli-$Z$ readouts are signed linear combinations of these probabilities. The measured map can therefore parameterize interactions over $O(d^2)$ coordinate pairs through a small set of shared circuit parameters, where $d$ is the bottleneck dimension. It provides a structured parameterization within, rather than beyond, the classical normalized-quadratic feature class. In a disjoint same-pipeline evaluation over eight GLUE-derived controlled interaction tasks and five shared seeds, SQUARE achieves an average test accuracy of $0.7565$, compared with $0.7355$ for an affine normalized-quadratic predictor, $0.7271$ for the evaluated parameter-matched Givens mixing model, $0.6817$ for an MLP, and $0.6155$ for a frozen-circuit control. Under reduced supervision, it also shows consistent gains over the strongest evaluated classical comparator, with the same qualitative pattern across multiple frozen LM backbones. All circuit experiments use simulation, while the learned feature map can be evaluated exactly in batched PyTorch without quantum hardware. These results indicate that how interaction coefficients are parameterized, rather than merely how many are learned, is a key design axis for compact adaptation over frozen language representations.
\end{abstract}

\section{Introduction}
Frozen language models (LMs) provide reusable representations for classification, scoring, reranking, and preference modeling without updating the backbone~\citep{NeurIPS2023Represent, NeurIPS2022what}. In this setting, a compact downstream module receives a fixed low-dimensional bottleneck $z\in\mathbb{R}^{d}$ and must make the target-relevant structure accessible to a lightweight task head. The central difficulty is not whether the frozen LM can encode rich information, but how a small post-bottleneck map should parameterize dependencies among its coordinates.

\begin{figure}[t]
\centering
\includegraphics[width=.98\linewidth]{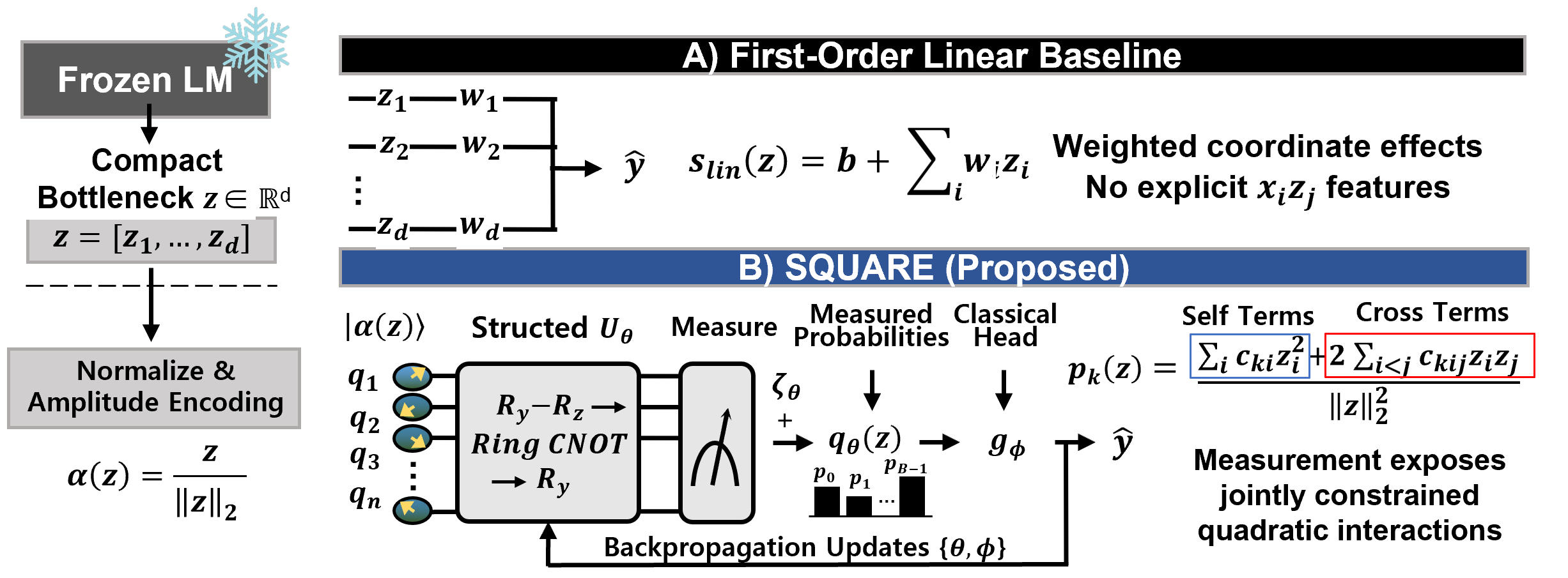}
\caption{
\textbf{First-order adaptation and SQUARE over frozen LM bottleneck.}
(A) A linear baseline aggregates weighted coordinate effects but does not explicitly output pairwise products. (B) SQUARE normalizes and amplitude-encodes $z$, learns structured circuit mixing, and measures a readout containing normalized $z_i^2$ and $z_i z_j$ terms. During training, simulator backpropagation jointly updates the circuit and classical head. For deployment, the learned map is compiled exactly into a batched PyTorch implementation with the same outputs. Shared circuit angles constrain many interaction coefficients, while inference remains fully compatible with classical frozen-LM pipelines.
}
\label{fig:1}
\vspace{-3mm}
\end{figure}

As illustrated in Fig.~\ref{fig:1}A, a first-order linear map aggregates weighted coordinate effects, but does not explicitly expose products $z_i z_j$ at its output. Bias-only and low-rank linear transformations similarly remain first order at this adaptation stage~\citep{ICLR2022PETL, ACL2022UniPELT, ICLR2022lora, ICML2024DoRA, ICLR2024_154926e0}. Interaction structure can instead be represented classically through explicit polynomial expansions, bilinear heads, factorization machines, kernels, and random features~\citep{rendle2010fm, rahimi2007rff, williams2001nystrom}; these alternatives differ primarily in how they constrain and share second-order coefficients.

We study the same design question through a parameterized quantum circuit: \emph{what coefficient structure is imposed when a compact frozen-LM bottleneck is amplitude-encoded, transformed, and measured?} SQUARE, summarized in Fig.~\ref{fig:1}B, applies a trainable $RY$--$RZ$--Ring-CNOT--$RY$ unitary and measures the result. One shared unitary generates the complete probability feature bank, so its low-rank PSD coefficient matrices are jointly compatible with the same complex rows and collectively resolve the identity. This coupled family---not quadraticity or entanglement alone---is the inductive bias evaluated here. We therefore center exact classical characterization and mechanism-matched comparison~\citep{NeurIPS2019inductive}.

We use \textbf{SQUARE}, a \textbf{S}tructured \textbf{QUA}ntum \textbf{RE}presentation adapter for frozen LM features, as a concrete case study of this question.
SQUARE freezes the LM, projects its representation into a low-dimensional bottleneck, and maps the bottleneck through the encoding--mixing--measurement pipeline in Fig.~\ref{fig:1}B~\citep{013091, 032430}.
The measured features are then passed to a lightweight classical head.
Measurement after a trainable unitary yields a structured nonlinear feature map: squaring linearly transformed amplitudes makes each basis probability a normalized quadratic form in the bottleneck coordinates~\citep{havlivcek2019supervised, liu2021rigorous, 062405, Nature2021power, PRX2021QNN}. These probability features are PSD and rank at most two and jointly resolve the identity; Pauli-$Z$ readouts are signed linear combinations of them and need not themselves be PSD or low rank. This gives SQUARE a two-stage interpretation: training uses a circuit-native encoding--unitary--measurement construction to learn the interaction structure, while deployment uses its exact algebraic image as a batched PyTorch module. The value of the quantum formulation is therefore the structured learning principle it induces, not a claim that the resulting function is inaccessible classically.

We evaluate SQUARE with controlled and generation-oriented protocols.
For controlled classification, we extract frozen LM representations from GLUE inputs and replace original labels with labels generated from nonlinear interaction rules~\citep{EMNLP2018GLUE, EMNLP2019design, CL2022PC}.
This preserves realistic language inputs while fixing the target dependency.
Because prediction over a frozen bottleneck is naturally an embedding-space problem, the comparison set is mechanism-matched, spanning classical second-order, kernel, non-parametric, graph-based, and manifold-based approximators (Section~\ref{sec:experiments}); we also report performance on the original, unmodified GLUE labels.
We further use candidate reranking, circuit ablations, and simulated noise sensitivity as scope-limited diagnostics~\citep{NeurIPS2022train, NeurIPS2022fewshot}. 

All quantum-circuit components are evaluated using a PennyLane-based simulator~\citep{miessen2024benchmarking, NeurIPS2024pqcapprox, NeurIPS2022VQC}. The simulator is an implementation of the circuit parameterization, not an algorithmic requirement: the same angles and gate product can be differentiated directly in PyTorch. After optimization, the learned unitary is compiled into batched complex-valued PyTorch, providing numerically equivalent classical inference without per-sample simulator calls or quantum hardware. Compact here refers to the number of trainable post-bottleneck parameters, not to a guaranteed training-time, memory, or latency advantage over specialized classical adapters.
Our primary experiment is the disjoint same-pipeline comparison in Section~\ref{sec:circuit_bias}: SQUARE reaches $0.7565$, compared with $0.7355$ for an affine normalized-quadratic score, $0.7271$ for the evaluated parameter-matched Givens-12 mixing model, $0.6817$ for the MLP, and $0.6155$ for a frozen circuit. Here and throughout, ``transfer across backbones'' means that the adapter protocol is retrained on regenerated controlled labels for each frozen backbone; it does not mean zero-shot transfer of one trained adapter or preservation of one semantic labeling rule.

Our contributions can be summarized as follows.
\textbf{(i) Jointly compatible quadratic feature parameterization.}
We establish that one common unitary generates the normalized rank-at-most-two PSD probability forms, distinguish their signed Pauli-$Z$ combinations, and characterize the circuit's locally low-dimensional image (Section~\ref{sec:square_feature_map}). The known amplitude-squaring identity is the starting point; the coupled feature family is the object studied here.
\textbf{(ii) Circuit-specified training with exact classical realization.}
We optimize the encoding--unitary--measurement map end-to-end and realize the same learned transformation as a numerically matched batched PyTorch module. The circuit serves as a compact parameterization language rather than a requirement for simulator-based training or inference.
\textbf{(iii) Mechanism-matched empirical validation.}
Controlled interaction tasks, compact-budget comparisons, and disjoint held-out controls show that the learned unitary-measurement constraint improves over the evaluated linear, MLP-style, frozen-circuit, and matched compact quadratic alternatives under the corresponding protocols. The result is evidence for a useful structured inductive bias in these controlled settings, not a claim of quantum computational advantage or universal superiority on natural-label NLP tasks.

\section{Related Work}

\BfPara{Parameter-Efficient Adaptation over Language Models}
Parameter-efficient adaptation updates only a small subset of LM parameters instead of the full pretrained backbone~\citep{ICLR2022PETL, ACL2022UniPELT}.
Representative methods include BitFit~\citep{ACL2022bitfit}, bottleneck adapters~\citep{ICML2019_Houlsby, NeurIPS2021_Compacter}, prefix- and prompt-tuning~\citep{ACL2021_PrefixTuning, EMNLP2021_PromptTuning}, low-rank methods such as LoRA, AdaLoRA, and DoRA~\citep{ICLR2022lora, ICLR2023adalora, ICML2024DoRA}, and orthogonal, low-precision, sparse, and more expressive low-rank variants~\citep{EMNLP2023Ortho, NeurIPS2023QLoRA, ICML2024_RoSA, ICLR2024_154926e0, ICML2024asymmetry, ICML2024LoRA+, NeurIPS2024Pissa, ICLR2026BoRA, ICLR2024increase}.
These methods are highly effective, but many compact variants primarily implement bias shifts, bottleneck transformations, or low-rank linear updates.
Our work studies a complementary setting in which Transformer weights are never adapted: all learning happens downstream of a fixed frozen bottleneck. Transformer-weight adaptation methods are therefore broad references rather than the mechanism-matched comparison set, and the LoRA- and adapter-style modules evaluated in our experiments are PEFT-inspired bottleneck transformations operating at the same stage as SQUARE.

\BfPara{Classical Interaction Models over Fixed Representations}
Learning over a fixed embedding has strong classical second-order solutions: polynomial expansions, bilinear heads, factorization machines~\citep{rendle2010fm}, kernel approximations~\citep{rahimi2007rff, williams2001nystrom}, and factorized polynomial models~\citep{chrysos2020pnets}. SQUARE belongs to this broader design space after compilation. Its narrower distinction is that one shallow complex-unitary circuit jointly generates a normalized probability feature bank. Amplitude-encoded feature maps are already known to correspond to low-degree polynomial kernels~\citep{havlivcek2019supervised, 032430}; our contribution is the analysis and controlled empirical test of this particular coupled restriction, not quadraticity or coefficient sharing in general.

\BfPara{Quantum-Enhanced Adaptation in Language Models}
Quantum-enhanced learning methods have recently been explored as compact hybrid modules inside large neural systems~\citep{NeurIPS2021pqpolicy, ICML2024qiren, ICLR2025design}.
For LM adaptation, QAA amplitude-encodes hidden states and merges quantum-transformed activations with frozen representations~\citep{CIKM2025QAA}, while QPA uses a parameterized quantum circuit (PQC) to generate PEFT weights~\citep{ICLR2025QPA}; both share a hybrid protocol in which the backbone remains classical and the quantum component is simulated~\citep{NeurIPS2024pqcapprox, NeurIPS2022VQC, miessen2024benchmarking, preskill2018quantum, kim2023evidence}.

SQUARE differs in its adaptation target: rather than encoding the full hidden state as in QAA or generating PEFT weights as in QPA, it operates on a compact projected subspace of frozen LM features, exposing measured interaction features for a downstream head~\citep{013091, 032430, havlivcek2019supervised, 062405, Nature2021power, PRX2021QNN}.
Table~\ref{tab:rel} and additional quantum machine learning (QML) background are provided in Appendix~\ref{app:rel}.

Our identity is the amplitude-encoding case of the kernel view of \citet{schuld2021kernel}; QuIC~\citep{raj2025quic} instead adapts Transformer weights with quantum-inspired orthogonal adapters.

\section{SQUARE}
\label{sec:square}
Let \(h=f_{\mathrm{LM}}(x)\in\mathbb{R}^{D}\) be a representation from a frozen language model, and let \(z=P(h)\in\mathbb{R}^{d}\) be a fixed compact bottleneck. Many compact adapters can be viewed as applying a restricted transformation before a task head,
%\[
$\hat{F}(z)=g_{\phi}(T_{\theta}z),$
%\]
where \(T_{\theta}\) denotes the adapter transformation and \(g_{\phi}\) is a lightweight task head with parameters \(\phi\).
These updates are parameter-efficient, but when instantiated as linear or low-rank transformations at the frozen-bottleneck stage, the adapter mapping itself remains linear in \(z\) and does not explicitly output pairwise products.
When the target depends on relations among features, such as \(z_i z_j\), comparisons, or compositions, the adapter must make such interaction structure accessible to the task head.
SQUARE instead explicitly constructs measured second-order features by amplitude-encoding \(z\), applying a parameterized unitary, and measuring the resulting state before the lightweight task head.

\subsection{Problem Formulation}
\label{sec:problem_formulation}
We formalize frozen-representation adaptation as learning a compact predictor over
%\[
$z=P(f_{\mathrm{LM}}(x))\in\mathbb{R}^{d},$
%\]
where neither the backbone \(f_{\mathrm{LM}}\) nor the projection \(P\) is updated. Only the adaptation module parameters \(\theta\) and task-head parameters \(\phi\) are trained.
We decompose the target rule over \(z\) into two components, i.e.,
%\begin{equation}
$F^\star(z)=F_{\mathrm{lin}}(z)+F_{\mathrm{int}}(z)$.
%\end{equation}
The first term captures independent evidence from individual coordinates,
%\begin{equation}
$F_{\mathrm{lin}}(z)=b+\sum_i a_i z_i$,
%\end{equation}
while the second captures pairwise and higher-order relations among coordinates,
%\begin{equation}
$F_{\mathrm{int}}(z)=\sum_{i<j}a_{ij}z_i z_j+\mathcal{H}(z)$.
%\end{equation}
Here, \(z_i z_j\) represents cross-feature interactions and \(\mathcal{H}(z)\) denotes task-specific nonlinear relations such as comparison, composition, or consistency. 
At the adapter stage, linear, bias-only, and low-rank transformations do not explicitly construct pairwise products of the bottleneck coordinates. SQUARE instead produces a measured feature map \(q_\theta(z)\) whose coordinates contain normalized cross-amplitude products, providing the downstream head with an explicit basis for second-order interaction structure.

The resulting predictor is
%\begin{equation}
$\hat{F}_{\theta,\phi}(z)
=
g_\phi(q_\theta(z))$
%\quad \text{or} \quad
or
$g_\phi([z;q_\theta(z)])$,
%\end{equation}
depending on whether the quantum-only or residual hybrid variant is used. This formulation is task-agnostic: \(\hat{F}_{\theta,\phi}(z)\) can be interpreted as a class logit, a scalar score, or a candidate-ranking score.

\subsection{SQUARE: Measured Quantum Feature Map}
\label{sec:square_feature_map}

SQUARE maps a compact frozen-LM representation into a measured quantum feature map. The module has three steps: encoding the representation into a quantum state, applying a trainable quantum circuit, and measuring the resulting state to obtain classical features for a lightweight task head.

\BfPara{Encoding}
Classical LM representations lie in a Euclidean feature space, while quantum circuits operate on state vectors in Hilbert space. SQUARE bridges these spaces by amplitude-encoding the compact frozen representation \(z\in\mathbb{R}^{d}\) into an \(n\)-qubit state, where \(n=\lceil \log_2 d\rceil\). Let \(B=2^n\). If \(d<B\), we zero-pad \(z\) to dimension \(B\); for notational simplicity, we denote the padded vector again by \(z\). Consequently, SQUARE prepares
\begin{equation}
|\psi(z)\rangle
=
\sum_{i=0}^{B-1}\nolimits
\alpha_i(z)|i\rangle,
\end{equation}
where $\alpha(z)=\frac{z}{\|z\|_2}$.
This is a representation change rather than a learned compression: the encoded inner product equals cosine similarity, so amplitude-encoding preserves the angular geometry of nonzero bottleneck directions but not norm magnitude. Nonlinear interaction features arise only after the circuit transformation and measurement (for the $\varepsilon$-stabilized normalization and the $z=0$ case: see Appendix~\ref{app:square_pennylane_code}).

\BfPara{Parameterized Quantum Circuit}
After encoding, SQUARE applies a parameterized quantum circuit \(U_\theta\) to
\(|\psi(z)\rangle\), whose trainable parameters \(\theta\) are rotation angles rather than weight matrices, so the circuit parameterizes a transformation of the encoded state before measurement; the parameters are optimized during training in the default SQUARE configuration.
Our default circuit uses a rotation--entanglement--rotation structure. Each layer first applies a trainable \(RYRZ\) rotation block to every qubit, then applies CNOT gates along a ring topology, and finally applies another trainable \(RY\) rotation to each qubit:
\begin{equation}
U_\theta
=
\prod_{\ell=1}^{L}
\left[
\left(
\prod_{q=1}^{n}
RY_q(\theta_{\ell q}^{(3)})
\right)
U_{\mathrm{ring}}
\left(
\prod_{q=1}^{n}
RZ_q(\theta_{\ell q}^{(2)})
RY_q(\theta_{\ell q}^{(1)})
\right)
\right].
\label{eq:square_pqc}
\end{equation}
Products are ordered by circuit application order (rightmost gates act first within each layer): the inner \(RYRZ\) block, the ring entanglement block, and a final \(RY\) refinement before measurement.

The ring operator is
\begin{equation}
U_{\mathrm{ring}}
=
\mathrm{CNOT}_{n,1}
\left(
\prod_{q=n-1}^{1}
\mathrm{CNOT}_{q,q+1}
\right).
\label{eq:ring_entanglement}
\end{equation}
The ring block connects all qubits with \(n\) CNOT gates per layer (one-indexed notation), a compact prespecified pattern relative to all-to-all entanglement; together, the rotations and topology determine the coefficient family exposed by measurement. Entanglement is not required for cross-coordinate products: product rotations can already create dense amplitude-coordinate rows whose squared magnitudes contain such terms. We therefore treat the ring as one architectural prior rather than as the source of interaction expressivity.

\BfPara{Quantum Measurement}
After processing via a parameterized circuit, the representation is still a quantum state.
SQUARE reads out classical features by concatenating two measurement readouts (the second is an exact linear function of the first; see below).
Let
%\begin{equation}
$|\varphi(z;\theta)\rangle = U_\theta|\psi(z)\rangle$
%\end{equation}
be the evolved quantum state, and let \(B=2^n\) denote the number of computational basis states.

First, SQUARE uses computational-basis probabilities:
\begin{equation}
p_j(z;\theta)
=
|\langle j|\varphi(z;\theta)\rangle|^2
=
|\langle j|U_\theta|\psi(z)\rangle|^2,
\qquad
j=0,\ldots,B-1.
\end{equation}
Second, it measures local Pauli-\(Z\) expectations:
\begin{equation}
\zeta_r(z;\theta)
=
\langle \varphi(z;\theta)|Z_r|\varphi(z;\theta)\rangle
=
\langle \psi(z)|U_\theta^\dagger Z_r U_\theta|\psi(z)\rangle,
\qquad
r=1,\ldots,n,
\end{equation}
where \(Z_r\) denotes the Pauli-\(Z\) operator acting on qubit \(r\).
The measured SQUARE feature is the concatenation of both readouts:
\begin{equation}
q_\theta(z)
=
[p_0(z;\theta),\ldots,p_{B-1}(z;\theta),
\zeta_1(z;\theta),\ldots,\zeta_n(z;\theta)].
\label{eq:square_readout}
\end{equation}
The readout dimension is \(B+n\), with no additional trainable readout parameters; \(n\) is small because SQUARE operates on a compact bottleneck, so the \(B=2^n\) probability readout remains tractable.

Both readouts can be written as expectation values of measurement operators. Collect the readout operators into
%\[
$\mathcal{M}
=
\{|j\rangle\langle j|\}_{j=0}^{B-1}
\cup
\{Z_r\}_{r=1}^{n}$.
%\]
For any \(M_k\in\mathcal{M}\), the corresponding measured coordinate is
\begin{equation}
\mu_k(z;\theta)
=
\langle \psi(z)|
U_\theta^\dagger M_k U_\theta
|\psi(z)\rangle .
\label{eq:square_measurement}
\end{equation}
This common form allows both probability features and Pauli-\(Z\) expectation features to be analyzed through the same measurement-induced interaction expansion.

\BfPara{Measurement-Induced Interactions}
SQUARE obtains nonlinear features because measurement after a trainable quantum
evolution is a quadratic form in the encoded amplitudes. Let \(z\neq 0\),
\(\alpha(z)=z/\|z\|_2\), and
%\[
$|\psi(z)\rangle=\sum_i \alpha_i(z)|i\rangle$.
%\]
For a measurement operator \(M_k\), the circuit \(U_\theta\) induces an effective
measurement operator
%\[
$A_k(\theta)=U_\theta^\dagger M_k U_\theta$.
%\]
Substituting this into Eq.~\eqref{eq:square_measurement} yields
%\[
$\mu_k(z;\theta)
=
\alpha(z)^\dagger A_k(\theta)\alpha(z)$.
%\]
Since \(M_k\) is Hermitian and \(U_\theta\) is unitary, \(A_k(\theta)\) is Hermitian. Thus, real-valued amplitude encoding becomes
\begin{equation}
\mu_k(z;\theta)
=
\sum_{i=0}^{B-1}\nolimits A_{k,ii}(\theta)\alpha_i(z)^2
+
2\sum_{0\le i<j<B}\nolimits
\mathrm{Re}(A_{k,ij}(\theta))\alpha_i(z)\alpha_j(z).
\label{eq:square_interaction_expansion}
\end{equation}
Because \(\alpha_i(z)\alpha_j(z)=z_i z_j/\|z\|_2^2\) for non-padded coordinates, the off-diagonal terms correspond to normalized cross-feature products. Therefore, the measured coordinates explicitly expose normalized cross-feature relations of the form \(z_i z_j/\|z\|_2^2\), providing the task head with features aligned with the pairwise interaction terms in \(F_{\mathrm{int}}(z)\).

\BfPara{Exact Quadratic Containment and Effective Readout Rank}
Because \(U_\theta\) is independent of \(z\), Eq.~\eqref{eq:square_interaction_expansion} is exact for the probability feature map: numerical reconstruction gives \(R^2=1.0000\) and maximum residual below \(5\times10^{-7}\). The readout is linearly redundant (\(\zeta_r=\sum_j(-1)^{j_r}p_j\), \(\sum_jp_j=1\)), so the nominal \((B{+}n)\)-dimensional readout at \(d=B=16\) has centered rank at most \(15\). Each real coefficient matrix $Q_j=a_ja_j^\top+b_jb_j^\top\succeq0$ has rank at most two and the matrices sum to identity. Jacobian ranks provide only local image dimensions (12 at sampled generic points for the default depth-1 circuit), not a global manifold claim. Real circuits form a restricted orthogonal-square family; the phased default circuit instead forms a restricted complex-unitary magnitude-square family. Exact quadraticity applies to $p_\theta(z)$, not automatically to a nonlinear head or to the separate boundary pipeline, which uses a learned nonlinear projection and $\sqrt p$ magnitude features (Appendix~\ref{app:structure_invariance}).

\subsection{Training and Parameterization} \label{subsec:training_param}
\BfPara{Hybrid Quantum-Classical Training}
Training follows a hybrid quantum-classical optimization procedure: the classical head parameters \(\phi\) are updated by standard backpropagation, and the circuit rotation angles \(\theta\) receive exact reverse-mode (backprop) gradients through the statevector simulation, composed with the head by the chain rule, so that \(q_\theta(z)\) and the task head are trained end-to-end while the LM backbone remains frozen (parameter shift is the hardware alternative; Appendix~\ref{app:training_details}).

\BfPara{Parameter Count}
All reported parameter counts include only trainable adapter and task-head parameters. The frozen LM \(f_{\mathrm{LM}}\) and fixed projection \(P\) are excluded. For SQUARE, the trainable quantum parameters are the circuit rotation angles. If \(d\) is not a power of two, we pad to the next power of two and use \(n=\lceil \log_2 d\rceil\) qubits. With depth \(L\) and three trainable rotations per qubit per layer, the number of quantum parameters is
$|\theta_q|=3Ln.$
The ring entanglement uses \(Ln\) CNOT gates and introduces no trainable parameters. Thus, the total number of trainable SQUARE parameters is
%\[
$|\theta_q|+|\phi|,$
%\]
where \(|\phi|\) denotes the parameters of the lightweight task head.
With \(g\) stored trainable rotations per qubit per layer, measurement width \(m\), \(C\) output logits, and an optional mixer with \(N_{\mathrm{mix}}\) parameters, the trainable post-bottleneck count is \(N_{\mathrm{SQUARE}} = gL\lceil \log_2 d\rceil + N_{\mathrm{mix}} + N_{\mathrm{head}}\), which for fixed \(d\), \(m\), and \(P\) does not depend on backbone width or depth. The primary held-out comparison uses the probability-only readout and a biased \(\mathrm{Lin}(16{\to}8)\)--\(\tanh\)--\(\mathrm{Lin}(8{\to}1)\) head, giving \(12+145=157\) trainable parameters; the separate validation/backbone suites use a 20-coordinate bias-free head and total 180. Adjacent $RY$ blocks across repeated layers can fuse exactly, so stored angles, an equivalent fused parameterization, and an observed local Jacobian rank are distinct quantities (Appendix~\ref{app:structure_invariance}). These are post-bottleneck parameter counts, not end-to-end computational-efficiency claims: backbone cost and simulator/runtime cost are reported separately. Appendix~\ref{app:param_accounting} gives the architecture and decomposition of the central configurations.

\BfPara{Simulator Training, Classical Inference}
Training and evaluation share one feature map through two implementations. The reported training runs execute \(q_\theta\) on a PennyLane statevector simulator, whose exact reverse-mode gradients compose with the autograd gradients of the head. The same parameterization is also differentiable in native PyTorch; Appendix~\ref{app:compute_resource} verifies agreement of both features and angle gradients and reports their runtime. For evaluation, we assemble \(U(\theta)\) as the same ordered gate product and compute \(|U(\theta)\psi(z)|^2\) in batched complex-valued PyTorch, with no per-sample simulator calls or sampling. The two inference paths agree to \(1.118\times10^{-7}\). Thus, the circuit specifies the architecture but neither simulator training nor quantum hardware is required to realize the learned model.

\section{Experiments}
\label{sec:experiments}
Our experiments test a deliberately scoped hypothesis: whether the coupled coefficient family induced by SQUARE is useful when labels depend on interactions over fixed LM bottleneck representations. The controlled labels are therefore a mechanism stress test, not evidence of broad semantic NLP superiority; original-label and reranking results are reported separately to delimit that scope.
We compare SQUARE with (i) mechanism-matched classical interaction models, including normalized quadratic, bilinear, factorization, Fourier, kernel, non-parametric, graph-based, and manifold methods; (ii) PEFT-inspired bottleneck transformations and MLP heads; and (iii) quantum-enhanced baselines QAA and QPA where applicable. Candidate configurations are selected using validation data only, and results are compared within each protocol. By holding the upstream representation fixed and using validation-only model selection, these protocols isolate differences in post-bottleneck parameterization rather than differences in backbone adaptation or representation learning.
Unless otherwise stated, all methods operate on the same frozen BERT bottleneck; for trainable adapter models, only post-bottleneck parameters are optimized. Throughout the paper, \emph{SQUARE} denotes the depth-1 $RYRZ$--Ring--$RY$ circuit over a $d{=}16$ bottleneck amplitude-encoded on $n{=}4$ qubits; protocol-specific readouts and lightweight heads are stated explicitly and account for the different parameter totals summarized in Appendix~\ref{app:param_accounting}. Additional frozen-backbone results with OPT-350M and GPT-2, as well as larger OpenLLaMA-3B and Mistral-7B backbones, are reported in Appendix~\ref{app:exp}. Results are averaged over three random seeds unless a table specifies otherwise; dataset construction, splits, metrics, and hyperparameters are detailed in Appendix~\ref{app:setup}. The main evidence is the disjoint same-pipeline comparison of circuit-induced and classical quadratic parameterizations in Section~\ref{sec:circuit_bias}; controlled boundary, original-label, backbone, and reranking results are complementary diagnostics reported with their protocol-specific scope.

\subsection{Diagnostic Nonlinear Decision Boundary Analysis}\label{sec:boundary}
\label{subsec:boundary}
\begin{figure}[t]
\centering
\includegraphics[width=.99\linewidth]{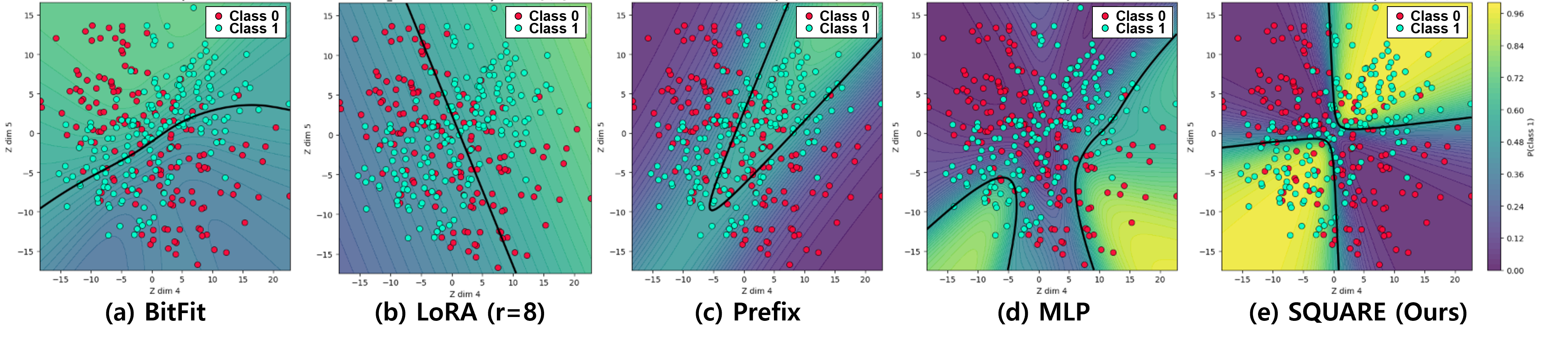}
\caption{
Learned decision surfaces on MNLI-derived two-dimensional PCA bottlenecks with controlled nonlinear labels.
}
\label{fig:2}
\vspace{-4mm}
\end{figure}

\begin{wraptable}{r}{0.5\linewidth}
\centering
\scriptsize
\caption{
Validation performance on the MNLI-derived two-dimensional PCA boundary task.
}
\label{tab:adapter_val_results}
\setlength{\tabcolsep}{3.5pt}
\renewcommand{\arraystretch}{1.05}
\begin{tabular}{l|c|ccc}
\toprule
Method & Params & Acc. & F1 & AUC \\
\midrule
BitFit & 33 & 0.6502 & 0.6463 & 0.6641 \\
LoRA$_{r{=}4}$ & 104 & 0.6383&  0.6283& 0.6536 \\
LoRA$_{r{=}8}$ & 200 & 0.6444 & 0.6378 & 0.6576 \\
Prefix & 289 & 0.6595 & 0.6562 & 0.6734 \\
MLP & 776 & 0.6869 & 0.6707 & 0.7217 \\
\textbf{SQUARE (Ours)} & 180 & \textbf{0.7416} & \textbf{0.7536} & \textbf{0.8259} \\
\bottomrule
\end{tabular}
\end{wraptable}

We use MNLI inputs only to obtain frozen BERT representations and discard the original MNLI labels.
The frozen representations are projected onto two principal components using a fixed PCA projection fitted on the training split, yielding a shared bottleneck \(z=(z_1,z_2)\) for all methods.
Binary labels are generated as
$y=\mathbf{1}\{f(z_1,z_2)>\tau\}$,
where $f$ contains a cross-coordinate interaction term and $\tau$ balances the training labels.
All methods receive the same PCA features and train only their adapter and task head; parameter counts include trainable adapter and task-head parameters only. 

Fig.~\ref{fig:2} visualizes the learned decision surfaces over the shared PCA bottleneck.
BitFit and LoRA produce relatively smooth boundaries and miss several curved regions of the controlled label structure.
Prefix and MLP yield more flexible surfaces, but still show visible mismatch with the imposed nonlinear boundary.
In contrast, SQUARE more closely follows the interaction-dependent regions, consistent with its measured feature map providing cross-coordinate features to the task head.
The black contour indicates the learned \(p(y=1 \mid z)=0.5\) boundary.

SQUARE achieves the best Accuracy, F1, and AUC with 180 trainable parameters (AUC \(0.8259\) vs.\ \(0.7217\) for MLP and \(0.6734\) for Prefix), indicating better recovery of this imposed interaction rule under the evaluated compact budget. Because the boundary pipeline includes a learned nonlinear projection and serves primarily as a visualization, it is not used to establish the main coefficient-structure claim; that claim is tested by the disjoint same-pipeline controls in Section~\ref{sec:circuit_bias}.

\BfPara{Boundary-Specific Comparison with Explicit Nonlinear Maps}
A natural question is whether this gap reflects the omission of explicit interaction features. We therefore evaluate two controlled geometries (checkerboard and radial ring) under identical splits, six seeds, and validation-only selection (Table~\ref{tab:boundary_lowsup}; per-geometry values in Appendix~\ref{app:boundary_details}). SQUARE reaches an average test AUC of \(0.992\), while classical Fourier (\(0.980\)) and explicit polynomial (\(0.975\)) maps are close behind and the evaluated GELU MLP reaches \(0.598\). We treat the MLP result as specific to the stated optimization protocol, not as evidence of representational inability; the diagnostic mainly confirms the value of explicit interaction features. Because the checkerboard rule lies outside the degree-two family, this experiment tests composition by the nonlinear head rather than membership of the target in SQUARE's feature class.
Two mechanism controls govern the interpretation (a broader approximator suite is in Appendix~\ref{app:broad_family}): freezing the circuit at random initialization changes boundary AUC by at most $0.022$ (Appendix~\ref{app:boundary_controls}), whereas on the disjoint held-out GLUE-derived suite it reduces mean accuracy from $0.7565$ to $0.6155$ (Appendix~\ref{app:heldout_validation}). Thus, structured random quadratic features can suffice when the head also receives $z$, while learning the circuit angles is important in the pure measured-feature path; Appendix~\ref{app:structure_invariance} gives the random-feature interpretation.

\begin{table}[t]
\centering
\scriptsize
\setlength{\tabcolsep}{3pt}\renewcommand{\arraystretch}{0.95}
\caption{Controlled interaction recovery on checkerboard and radial-ring tasks. Boundary is the full-supervision mean over both tasks; reduced-supervision results report mean test AUC $\pm$ standard deviation over six seeds. Full results are in Appendix~\ref{app:boundary_details}.}
\label{tab:boundary_lowsup}
\begin{tabular}{l|c|c|cccc}
\toprule
Method & Params & Boundary & 10\% & 25\% & 50\% & 100\% \\
\midrule
Classical Fourier & 210 & 0.980 & 0.736 $\pm$ 0.125 & 0.830 $\pm$ 0.160 & 0.879 $\pm$ 0.156 & \textbf{0.981 $\pm$ 0.020} \\
Explicit polynomial & 386 & 0.975 & 0.688 $\pm$ 0.049 & 0.851 $\pm$ 0.083 & 0.820 $\pm$ 0.167 & 0.969 $\pm$ 0.019 \\
MLP & 961 & 0.598 & 0.515 $\pm$ 0.026 & 0.599 $\pm$ 0.027 & 0.563 $\pm$ 0.092 & 0.616 $\pm$ 0.122 \\
\textbf{SQUARE (Ours)} & 281 & \textbf{0.992} & \textbf{0.823 $\pm$ 0.098} & \textbf{0.911 $\pm$ 0.052} & \textbf{0.948 $\pm$ 0.026} & 0.979 $\pm$ 0.016 \\
\bottomrule
\end{tabular}
\end{table}

\subsection{GLUE-Derived Controlled Interaction Classification: Circuit-Induced Quadratic Bias vs.\ Classical Parameterizations}
\label{subsec:glue_controlled}
\label{sec:circuit_bias}
\vspace{-2mm}

\BfPara{Controlled Interaction Protocol}
We use inputs from eight GLUE datasets as distinct frozen-representation distributions while replacing their original semantic labels with one prespecified interaction-rich rule over training-fitted PCA bottlenecks. The benchmark therefore evaluates recovery of the same controlled mechanism across eight realistic language-input distributions rather than eight independent natural-language targets. For the primary comparison, we construct disjoint train, validation, and test subsets for every task. The feature scaler, PCA projection, and median label threshold are fitted using training data only, checkpoints are selected by validation AUC, and test sets are evaluated once. All methods receive the same frozen bottleneck features and are evaluated over five shared seeds.

\begin{table}[t]
\vspace{-2mm}
\centering
\scriptsize
\caption{\textbf{Core held-out evidence.} Gap is SQUARE minus the comparator in eight-task average test accuracy. Complete scores, parameter counts, and uncertainty summaries are in Appendix~\ref{app:heldout_validation}.}
\label{tab:heldout_evidence_summary}
\setlength{\tabcolsep}{5pt}
\renewcommand{\arraystretch}{1.00}
\begin{tabular}{lcl}
\toprule
Comparator & Gap & Primary contrast \\
\midrule
Frozen circuit & $+0.141$ & Trained vs. fixed mixing \\
Head only & $+0.115$ & Measured vs. no measured map \\
MLP & $+0.075$ & Interaction-feature architecture \\
Givens-12 & $+0.029$ & Matched-angle mixing family \\
Norm.\ quadratic & $+0.021$ & Coupled vs. affine quadratic score \\
\midrule
\multicolumn{3}{l}{SQUARE avg.\ test accuracy: \textbf{0.7565}} \\
\bottomrule
\end{tabular}
\vspace{-3mm}
\end{table}

\BfPara{Matched Quadratic Parameterizations}
The experiment asks whether SQUARE benefits merely from normalized quadratic lifting or from the particular coefficient constraints induced by its circuit. The head-only control removes the measured feature map, whereas the frozen-circuit control retains fixed quadratic features without learning the circuit angles. The normalized-quadratic score provides a flexible explicit second-order predictor. Givens-12 uses the same 12 learned mixing parameters as SQUARE, followed by squared features and the same nonlinear head, making it the closest parameter- and mechanism-matched comparison evaluated here. This comparison isolates the implemented 12-rotation schedule, not all structured classical mixing families: a sparse plane-rotation product and a parameter-shared tensor-product circuit can differ in coordinate connectivity despite equal parameter counts. Table~\ref{tab:heldout_evidence_summary} summarizes the attribution tests; complete absolute scores, parameter counts, and taskwise interval summaries are reported in Appendix~\ref{app:heldout_validation}.

\BfPara{Held-Out Results}
SQUARE obtains the highest average test accuracy among the evaluated same-pipeline models (Table~\ref{tab:heldout_evidence_summary}). The head-only and frozen-circuit comparisons show that neither the downstream head nor fixed quadratic lifting alone explains the result. The cleanest structural comparison is Givens-12: despite using the same number of learned mixing parameters, squared features, and nonlinear head, it trails SQUARE by $0.029$ average accuracy. SQUARE also exceeds the normalized-quadratic score by $0.021$, although this comparison additionally differs in head composition because the latter is an affine scalar predictor.

Together, these comparisons support a focused conclusion: under this controlled held-out protocol, the jointly constrained unitary-measurement parameterization provides a useful inductive bias within the classical normalized-quadratic family. The result does not establish universal superiority over quadratic models or quantum computational advantage. The topology diagnostic in Appendix~\ref{app:corrected_ablation} further shows that the prespecified $RYRZ$--ring circuit is the strongest observed configuration on three held-out distributions, while the corresponding $RY$--ring variant is weaker than its non-entangling counterpart. We therefore interpret performance as a property of the jointly learned rotation--topology design rather than of entanglement in isolation. Compact-budget factorization results and complete protocol details are reported in Appendices~\ref{app:glue_extra} and~\ref{app:heldout_validation}.

\BfPara{Why the Constraint Matters}
An unrestricted quadratic predictor assigns its coefficients independently, whereas SQUARE generates a collection of low-rank positive-semidefinite forms from one shared unitary and a common measurement basis. The circuit angles therefore change many interaction coefficients jointly rather than estimating every pair in isolation. The held-out ordering is consistent with---but does not by itself prove---a finite-sample regularization effect: SQUARE improves on the affine normalized-quadratic score and on the evaluated Givens-12 construction. The latter is the cleaner matched-head comparison; the former also changes head composition. This is the central empirical claim of the paper, limited to the implemented controls and controlled target family. We do not infer a generalization theorem or superiority over untested Householder, butterfly, Cayley, or dense-base orthogonal parameterizations.

\BfPara{Robustness and Deployment Scope}
The same qualitative SQUARE--MLP ordering appears when the controlled-label protocol is rebuilt over OPT-350M, GPT-2, OpenLLaMA-3B, and Mistral-7B representations, with within-protocol gains of $0.0334$--$0.0662$ (Appendix~\ref{app:exp}). These experiments retrain the adapter for each representation and therefore test robustness across frozen feature geometries rather than zero-shot transfer. Once training is complete, the learned unitary can be materialized as the identical batched PyTorch map described in Section~\ref{subsec:training_param}; deployment consequently preserves the learned coefficient structure without simulator calls or quantum hardware.

\BfPara{Original-Label GLUE Performance}
On the original, unmodified labels of SST-2, RTE, MRPC, and QNLI (Table~\ref{tab:original_glue}, Appendix~\ref{app:glue_extra}), SQUARE scores \(0.686\) with \(49\) parameters, versus \(0.686\) with \(5{,}041\) for the MLP, \(0.673\) with \(445\) for the explicit quadratic head, and \(0.755\) for encoder-adapted LoRA with \(221{,}953\). Original-label GLUE therefore supports trainable-parameter compactness rather than an accuracy advantage; the strong gains are tied to the controlled interaction-label protocols and should not be generalized to natural-label classification.

\section{Conclusion}
We introduced SQUARE as a circuit-induced parameterization of normalized quadratic features over frozen LM bottlenecks. A shared unitary jointly constrains its PSD probability features, providing a compact coefficient-sharing structure for learning second-order interactions. Controlled experiments show improvements over frozen mixing, the evaluated parameter-matched Givens construction, and an affine normalized-quadratic score, with consistent qualitative behavior across multiple frozen LM backbones. The learned map can also be realized exactly in batched PyTorch without quantum hardware.
These results position SQUARE as a structured parameterization within the classical normalized-quadratic family, rather than as a uniquely quantum function class or a source of computational advantage. The current evidence is limited to low-dimensional controlled interaction
settings and does not establish broad gains on natural-label NLP tasks.
Broader validation under natural supervision, higher-dimensional bottlenecks,
coordinate-permuted protocols, and additional structured parameterizations
is therefore an important direction for future work.

% \section{Conclusion}
% We introduced SQUARE as a circuit-induced parameterization of normalized quadratic features over frozen LM bottlenecks. One shared unitary jointly constrains its complete PSD probability feature bank, and the central controlled experiment shows improvements over frozen mixing, the evaluated matched Givens construction, and an affine normalized-quadratic score. We claim neither a uniquely quantum function class nor computational advantage. Instead, the circuit is a compact parameterization language for a testable coefficient-sharing prior with an exact PyTorch realization. This establishes a scoped connection between common-unitary compatibility and controlled interaction learning; broader natural-task value and dimensional scaling remain open.

\subsubsection*{AI Use Statement}
Generative AI tools were used during manuscript preparation for language refinement, proofreading, and improving the clarity and organization of the presentation. The authors independently reviewed and verified all scientific claims, mathematical derivations, experimental designs, implementations, reported results, and references, and retain full responsibility for the content of the paper.

\subsubsection*{Reproducibility Statement}
Dataset construction, data splits, controlled label-generation rules, baseline families and their search spaces, model-selection protocols, evaluation metrics, and hyperparameters are specified in Appendices~\ref{app:setup} and~\ref{app:baseline_implementation}. The direct PyTorch realization of the learned SQUARE feature map is described operator-by-operator in Appendix~\ref{app:compute_resource}, including its numerical agreement with the PennyLane implementation. All quantum components are simulator-based, and the experimental pipeline can be reproduced without access to quantum hardware.

\subsubsection*{Ethics Statement}
Because SQUARE operates on frozen language-model representations, it may inherit biases, spurious correlations, or other limitations present in the underlying backbones. Downstream use should therefore include appropriate task-specific validation, fairness and reliability checks, privacy safeguards, and human oversight where applicable.

\bibliographystyle{square_preprint}
\bibliography{reference}

%%%%%%%%%%%%%%%%%%%%%%%%%%%%%%%%%%%%%%%%%%%%%%%%%%%%%%%%%%%%
\newpage
\appendix
\section*{APPENDIX}
\section{Background on Quantum Machine Learning}
This section provides a background on gate-based quantum machine learning used in SQUARE.
The goal is not to give a complete introduction to quantum computing, but to make the main ingredients of our adapter self-contained for readers who may be less familiar with quantum machine learning.
SQUARE relies on a standard hybrid quantum--classical workflow: a classical language-model representation is first encoded into a quantum state, transformed by a parameterized quantum circuit, and then converted back into classical features through measurement.
Understanding this pipeline requires a few basic concepts, including qubits, Hilbert-space representations, amplitude encoding, trainable quantum circuits, and measurement readouts.
We therefore review these components in the same order in which they appear in SQUARE, emphasizing how they relate to the measured quantum feature map used in the main paper.

\subsection{Qubits and Hilbert-Space Representations}
\label{app:qml_qubits}

A classical bit takes one of two values, \(0\) or \(1\).
A qubit is the quantum analogue of a bit and is represented as a normalized vector in a two-dimensional complex Hilbert space:
\begin{equation}
|\psi\rangle
=
\alpha |0\rangle + \beta |1\rangle,
\qquad
\alpha,\beta\in\mathbb{C},
\qquad
|\alpha|^2+|\beta|^2=1.
\end{equation}
Here, \(|0\rangle\) and \(|1\rangle\) are computational basis states, and \(\alpha,\beta\) are probability amplitudes.
When the qubit is measured in the computational basis, the probability of observing \(|0\rangle\) is \(|\alpha|^2\), and the probability of observing \(|1\rangle\) is \(|\beta|^2\).

For an \(n\)-qubit system, the Hilbert space has dimension \(2^n\).
Its computational basis is
\[
\{|0\rangle,\ldots,|2^n-1\rangle\},
\]
or equivalently the binary basis states
\[
\{|00\cdots 0\rangle,\ldots,|11\cdots 1\rangle\}.
\]
A general \(n\)-qubit pure state is
\begin{equation}
|\psi\rangle
=
\sum_{i=0}^{2^n-1}
\alpha_i |i\rangle,
\qquad
\sum_{i=0}^{2^n-1}
|\alpha_i|^2=1.
\end{equation}
This exponential state dimension is one reason quantum feature maps are attractive as compact representations.
However, only limited information can be extracted through measurement, so the design of the encoding, circuit, and readout is crucial.

\subsection{Parameterized Quantum Circuits}
\label{app:qml_pqc}
After encoding, a quantum circuit applies a sequence of unitary transformations to the input state.
A unitary transformation \(U\) satisfies
\[
U^\dagger U = UU^\dagger = I,
\]
so it preserves the norm of the quantum state.
In QML, the unitary transformation is usually parameterized by trainable angles \(\theta\), yielding a parameterized quantum circuit (PQC), also called a variational ansatz:
\begin{equation}
|\varphi(z;\theta)\rangle
=
U_\theta |\psi(z)\rangle.
\end{equation}
The parameters \(\theta\) are analogous to neural network weights, but they appear as rotation angles inside quantum gates.

Single-qubit rotation gates are common trainable operations.
For example, Pauli rotations around the \(y\)- and \(z\)-axes are
\begin{equation}
RY(\theta)
=
\begin{bmatrix}
\cos(\theta/2) & -\sin(\theta/2) \\
\sin(\theta/2) & \cos(\theta/2)
\end{bmatrix},
\qquad
RZ(\theta)
=
\begin{bmatrix}
e^{-i\theta/2} & 0 \\
0 & e^{i\theta/2}
\end{bmatrix}.
\end{equation}
Multi-qubit gates such as CNOT introduce dependencies across qubits.
When a circuit cannot be decomposed into independent operations on each qubit, it can create entangled states.
Entanglement is important for SQUARE because it allows the effective measurement operator to couple amplitudes associated with different basis states.

\begin{figure}[t]
\centering
\includegraphics[width=0.90\linewidth]{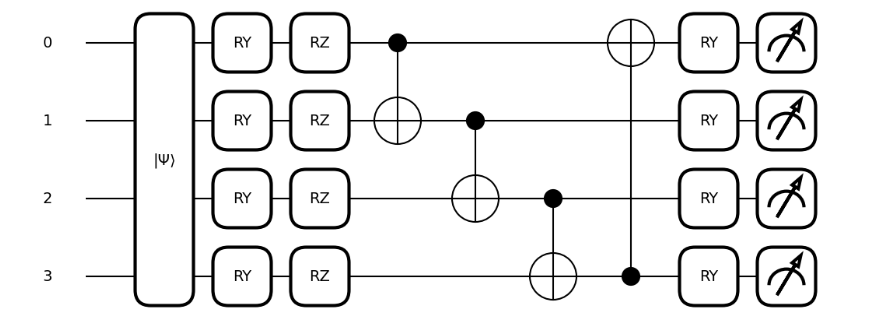}
\caption{
Visualization of the SQUARE parameterized quantum circuit used in the main experiments.
The input bottleneck representation is amplitude-encoded into an \(n\)-qubit state \(|\psi(z)\rangle\).
Each layer applies an \(RYRZ\) rotation block, ring CNOT entanglement, and a final \(RY\) rotation block before measurement.
The shown example uses \(n=4\) qubits and one circuit layer.
}
\label{fig:square_circuit}
\end{figure}

In the main method, SQUARE uses a compact rotation--entanglement--rotation layer.
Each layer first applies trainable \(RY\) and \(RZ\) rotations, then ring CNOT entanglement, and finally another trainable \(RY\) block.
For depth \(L\), this gives
\begin{equation}
U_\theta
=
\prod_{\ell=1}^{L}
\left[
\left(
\prod_{q=1}^{n}
RY_q(\theta_{\ell q}^{(3)})
\right)
U_{\mathrm{ring}}
\left(
\prod_{q=1}^{n}
RZ_q(\theta_{\ell q}^{(2)})
RY_q(\theta_{\ell q}^{(1)})
\right)
\right].
\end{equation}
The ring entanglement operator is
\begin{equation}
U_{\mathrm{ring}}
=
\mathrm{CNOT}_{n,1}
\left(
\prod_{q=n-1}^{1}
\mathrm{CNOT}_{q,q+1}
\right),
\end{equation}
where we use one-indexed qubit notation and, exactly as in Eq.~\eqref{eq:ring_entanglement}, products are ordered by application with the rightmost gate acting first: $\mathrm{CNOT}_{1,2}$ is applied first, then $\mathrm{CNOT}_{2,3},\ldots,\mathrm{CNOT}_{n-1,n}$, and finally $\mathrm{CNOT}_{n,1}$. This single canonical ring is the one implemented in Listing~\ref{lst:square_pennylane} and drawn in Fig.~\ref{fig:square_circuit}; for the basis input $|1000\rangle$ it produces $|0111\rangle$ (the opposite ordering would give $|1100\rangle$ and is not what the code implements).
This topology connects all qubits with \(n\) CNOT gates per layer, providing a compact interaction structure compared with all-to-all entanglement.
Fig.~\ref{fig:square_circuit} visualizes the complete SQUARE circuit used in the main experiments.
\subsection{Derivation of Measurement-induced Interaction Features}
\label{app:measurement_interaction_derivation}

We provide the derivation behind Eq.~\ref{eq:square_interaction_expansion}.
For notational clarity, first assume \(d=2^n\). Given a nonzero bottleneck
representation \(z\in\mathbb{R}^{d}\), amplitude encoding prepares
\[
|\psi(z)\rangle
=
\sum_{i=0}^{d-1}
\alpha_i(z)|i\rangle,
\qquad
\alpha_i(z)=\frac{z_i}{\|z\|_2}.
\]
Let \(M_k\) be a Hermitian measurement operator and \(U_\theta\) be the unitary
implemented by the parameterized quantum circuit. The measured coordinate is
\[
\mu_k(z;\theta)
=
\langle \psi(z)|U_\theta^\dagger M_k U_\theta|\psi(z)\rangle .
\]
Define the effective measurement operator
\[
A_k(\theta)=U_\theta^\dagger M_k U_\theta .
\]
Because \(M_k=M_k^\dagger\) and \(U_\theta^\dagger U_\theta=I\), we have
\[
A_k(\theta)^\dagger
=
\left(U_\theta^\dagger M_k U_\theta\right)^\dagger
=
U_\theta^\dagger M_k^\dagger U_\theta
=
U_\theta^\dagger M_k U_\theta
=
A_k(\theta),
\]
so \(A_k(\theta)\) is Hermitian. Substituting the amplitude-encoded state gives
\[
\mu_k(z;\theta)
=
\sum_{i=0}^{d-1}\sum_{j=0}^{d-1}
\alpha_i(z) A_{k,ij}(\theta)\alpha_j(z),
\]
where the amplitudes are real-valued under the encoding used in SQUARE.
Separating diagonal and off-diagonal terms yields
\[
\mu_k(z;\theta)
=
\sum_{i=0}^{d-1}
A_{k,ii}(\theta)\alpha_i(z)^2
+
\sum_{i<j}
\left(
A_{k,ij}(\theta)+A_{k,ji}(\theta)
\right)
\alpha_i(z)\alpha_j(z).
\]
Since \(A_k(\theta)\) is Hermitian,
\(A_{k,ji}(\theta)=A_{k,ij}(\theta)^*\). Therefore,
\[
A_{k,ij}(\theta)+A_{k,ji}(\theta)
=
2\mathrm{Re}\!\left(A_{k,ij}(\theta)\right),
\]
and
\[
\mu_k(z;\theta)
=
\sum_{i=0}^{d-1}
A_{k,ii}(\theta)\alpha_i(z)^2
+
2\sum_{0\le i<j<d}
\mathrm{Re}\!\left(A_{k,ij}(\theta)\right)
\alpha_i(z)\alpha_j(z).
\]
Using \(\alpha_i(z)=z_i/\|z\|_2\), we obtain
\[
\mu_k(z;\theta)
=
\sum_{i=0}^{d-1}
A_{k,ii}(\theta)\frac{z_i^2}{\|z\|_2^2}
+
2\sum_{0\le i<j<d}
\mathrm{Re}\!\left(A_{k,ij}(\theta)\right)
\frac{z_i z_j}{\|z\|_2^2}.
\]
Thus, measurement after trainable quantum evolution exposes both coordinate-wise
terms \(z_i^2\) and cross-coordinate interaction terms \(z_i z_j\).
If \(d\) is not a power of two, SQUARE zero-pads \(z\) to dimension
\(B=2^{\lceil \log_2 d\rceil}\) before normalization. The same derivation holds
with \(d\) replaced by \(B\). For padded coordinates, \(z_i=0\), so all terms
involving padded dimensions vanish. Therefore, zero-padding does not change the
interaction expansion over the original non-padded coordinates.

\greybox{
\begin{remark}[How SQUARE learns nonlinear features]
The coefficients of the interaction terms in
Eq.~\ref{eq:square_interaction_expansion} are not fixed. They are determined by
\[
A_k(\theta)=U_\theta^\dagger M_k U_\theta,
\]
which changes as the circuit angles \(\theta\) are optimized.
Trainable rotations adjust the measurement basis and determine which amplitude products are emphasized by the downstream loss.
The ring entanglement block changes the attainable coefficient family relative to product rotations. It is not required for cross-coordinate products, which can already arise when dense amplitude-coordinate rows produced by local rotations are squared; Appendix~\ref{app:corrected_ablation} therefore treats topology as a diagnostic design choice rather than attributing all interactions to entanglement.
Measurement then reads these learned cross-amplitude relations as classical coordinates \(\mu_k(z;\theta)\), which form the measured feature vector \(q_\theta(z)\) used by a lightweight head to model interaction-dependent components of the target.
\end{remark}
}

\subsection{Structural Characterization and Scale Invariance of the Measured Map}
\label{app:structure_invariance}
\BfPara{Structural Characterization of the Coefficient Matrices}
The quadratic identity itself is a direct consequence of squaring linearly transformed amplitudes, and closely related outer-product/kernel views of amplitude-encoded models are known~\citep{schuld2021kernel, havlivcek2019supervised}; the contribution we claim is therefore the specific restricted parameterization and the evidence for its usefulness, not the containment result. The following are necessary structural properties of every circuit-induced collection, not a sufficient characterization of all collections realizable by one common unitary. Write row $j$ of $U_\theta$ as $u_j=a_j+ib_j$ with $a_j,b_j\in\mathbb{R}^{B}$. For real inputs, $p_j(z)=\big((a_j^\top\alpha)^2+(b_j^\top\alpha)^2\big)$ with $\alpha=z/\|z\|_2$, so the real coefficient matrix of the $j$-th probability feature is
\begin{equation}
Q_j \;=\; a_j a_j^\top + b_j b_j^\top,\qquad Q_j\succeq 0,\qquad \operatorname{rank}(Q_j)\le 2,\qquad \sum_{j=0}^{B-1} Q_j = I,
\label{eq:psd_rank2}
\end{equation}
where the last identity follows from unitarity ($\sum_j u_j u_j^\dagger=I$). SQUARE is thus a jointly constrained collection of $B$ positive-semidefinite, rank-at-most-two normalized quadratic probability features whose coefficient matrices resolve the identity, followed by a task head; the Pauli-$Z$ features are signed sums of the $Q_j$ and add no new information. Joint compatibility of all rows with one unitary imposes further constraints beyond PSD, rank, trace, and resolution of identity. This characterization identifies the natural classical comparison point---structured linear projections followed by squared magnitudes---evaluated in Section~\ref{sec:circuit_bias}. Two distinctions should be kept explicit. First, potentially nonzero coefficients on $O(d^2)$ coordinate pairs are not $O(d^2)$ independently accessible features: the readout has centered rank at most $15$ at $d=B=16$. Second, ``normalized quadratic'' refers to $p_\theta(z)=z^\top Qz/(z^\top z)$; it is homogeneous of degree zero in the unnormalized coordinates, and composition with a nonlinear task head is not itself a quadratic predictor.

\BfPara{Scale Invariance and the Input Path}
Because $\alpha(cz)=\operatorname{sign}(c)\,\alpha(z)$ and every coordinate of $q_\theta$ is quadratic in $\alpha$, we have $q_\theta(cz)=q_\theta(z)$ for every real $c\neq0$. Any predictor of the form $g_\phi(q_\theta(z))$ is consequently invariant to input scaling and global sign and cannot distinguish two points on the same ray with different radii; in particular it cannot exactly represent a rule whose label depends on $\|z\|$. Section~\ref{sec:problem_formulation} therefore admits pure and residual paths, and we state which is used in each protocol. The GLUE-derived controlled tasks and additional-backbone experiments use the pure path on a $d{=}16$ PCA bottleneck; the Alpaca rebuild reports both variants. The boundary experiments instead map PCA-2 coordinates $x$ to $v=\tanh(Wx+b)\in\mathbb{R}^{16}$, amplitude-encode $v$, and feed the affine head $[v;\,\sqrt{p_\theta(v)}]\in\mathbb{R}^{32}$. Although $p_\theta(v)$ is normalized quadratic in $v$, $\sqrt{p_\theta(v)}=|U_\theta\alpha(v)|$ is a magnitude readout and $v$ is nonlinear in $x$; the end-to-end boundary model is therefore not exactly quadratic in $x$. The learned projection can itself encode radial information into direction, so these experiments do not establish that concatenating $v$ is mathematically necessary. Their component controls are interpreted only within this broader boundary architecture.

\BfPara{What Subset of Quadratic Models the Circuit Induces}
Eq.~\eqref{eq:psd_rank2} can be sharpened into a quantitative description of the family $\{Q_j(\theta)\}$. Because each row $u_j$ of a unitary has unit norm, $\operatorname{tr}Q_j=1$: every probability feature is PSD with rank at most two, and the $B$ forms resolve the identity. If the circuit is real, then $b_j=0$ and the model is a restricted orthogonal-projection-square family. With phases, the default circuit belongs instead to a restricted complex-unitary magnitude-square family, and the $RZ$ block can produce rank-two real coefficient matrices; the complex family is not a subfamily of the real orthogonal-square family. Smoothness bounds local image dimension by the angle count but does not establish a single global manifold. At sampled generic parameter points, the Jacobian rank of $\theta\mapsto(Q_j)_j$ was $12$, $20$, and $28$ for $L=1,2,3$ at $n=4$, and $8$ for the real $L=1$ circuit. We report these as observed local image dimensions, not globally identified degrees of freedom. Table~\ref{tab:structure_comparison} separately reports stored scalars and effective coefficient-family constraints. The ring does not make $Q_j$ sparse in the coordinate basis (each row of $U_\theta$ is generically dense), but it does tie the family to the coordinate ordering of the bottleneck: the attainable set is not closed under arbitrary input permutations, so coordinate ordering remains part of the parameterization.

\begin{table}[h]
\centering\footnotesize
\caption{Structural comparison at $d=16$. Stored scalars are implementation parameters before a head; the last column gives an effective-family bound or structural constraint.}
\label{tab:structure_comparison}
\resizebox{\linewidth}{!}{\begin{tabular}{llrl}
\toprule
Model & Coefficient-matrix structure & Stored & Effective family \\
\midrule
Full explicit quadratic & any symmetric $A$ & $136$ & $\le136$ \\
Low-rank bilinear (rank $r$) & $A=\operatorname{Sym}(UV^\top)$ & $32r$ & rank $\le2r$; dim. $\le\min(136,32r)$ \\
Factorization machine (rank $k$) & off-diagonal coefficients from $VV^\top$ & $16k$ & Gram-constrained off diagonal \\
Learned mixing + squares & $A=W\operatorname{diag}(c)W^\top$ & $256{+}16$ & $\le136$ \\
Learned orthogonal mixing + squares & spectral form, $W\in O(16)$ & $120{+}16$ & $\le136$ \\
SQUARE, real ($L=1$) & restricted orthogonal squares & $8{+}16$ & sampled local rank $\le8$ before $c$ \\
SQUARE, complex ($L$ layers) & restricted unitary magnitudes & $12L{+}16$ stored & $4{+}8L$ after exact adjacent-$RY$ fusion; sampled local rank no larger \\
\bottomrule
\end{tabular}}
\end{table}
The implementation stores $3Ln$ rotation angles. For $L>1$, the terminal $RY$ block of one layer and the initial $RY$ block of the next are adjacent and fuse exactly via $RY(a)RY(b)=RY(a+b)$. The same unitary therefore admits an equivalent parameterization with at most $3Ln-n(L-1)=n(2L+1)$ angles, giving $4+8L$ at $n=4$. Stored angles, this analytic gate-fusion bound, and observed local Jacobian rank are distinct quantities; the numerical ranks $12$, $20$, and $28$ match the bound at the sampled generic points but are not asserted as a global-dimension theorem.

\BfPara{The Random Circuit as a Structured Random-Feature Map}
The untrained-circuit result of Appendix~\ref{app:boundary_controls} (trained-minus-untrained AUC within $[-0.017,+0.022]$ on the residual-path boundary suite) invites a random-feature reading, which the structure above makes precise. For a Haar-random unitary $U$ and real unit vectors $\alpha,\alpha'$, $\mathbb{E}_U[Q_j]=I/B$ and $\mathbb{E}_U\big[\sum_j p_j(z)p_j(z')\big]=\big(1+(\alpha^\top\alpha')^2\big)/(B+1)$, so the expected linear kernel of the probability features is an affine function of squared cosine similarity---a normalized degree-two polynomial kernel---and a random circuit is a $B$-feature Monte Carlo approximation of it. For the default ring circuit with angles drawn uniformly from $[0,2\pi)$, a numerical study ($4{,}000$ draws, $50$ random pairs) finds $\mathbb{E}_\theta[Q_j]$ diagonal to within $0.002$, with diagonal entries $0.061$--$0.065$ (Haar: $1/16=0.0625$). Its expected kernel lies within $0.008$ of the Haar value but retains orientation dependence relative to the coordinate axes. The random SQUARE circuit is therefore an anisotropic, coordinate-aligned quadratic feature map. This interpretation is consistent with the residual-path boundary result, while the disjoint held-out pure-path suite shows that training the circuit improves mean accuracy from $0.6155$ to $0.7565$ (Appendix~\ref{app:heldout_validation}). On that same protocol, trained SQUARE also exceeds the normalized-quadratic and parameter-matched Givens-12 controls, indicating that the learned circuit constraint contributes beyond quadratic lifting alone under this evaluation.

\subsection{Hybrid Quantum-Classical Training Details}
\label{app:training_details}
Training follows a hybrid quantum-classical optimization procedure. The classical head parameters \(\phi\) are updated by standard backpropagation, while the quantum parameters \(\theta\) are trainable rotation angles in the PQC. \textbf{Differentiation method actually used.} In all reported experiments the circuit is executed on a statevector simulator and $\partial\mu_k/\partial\theta_r$ is obtained by reverse-mode automatic differentiation \emph{through the simulation}: the GLUE-derived, reranking, and additional-backbone experiments use PennyLane \texttt{default.qubit} with \texttt{interface="torch"} and \texttt{diff\_method="backprop"} (not the \texttt{"best"} default and not parameter shift), and the decision-boundary control suite uses a native PyTorch statevector implementation of the same gates, differentiated by PyTorch autograd. These gradients are exact for the simulated circuit; in a numerical check over 20 random draws, backpropagation and parameter shift agreed to a maximum absolute difference of $1.1\times10^{-15}$, and no gradient entry was zero. For Pauli rotation gates, if a measured coordinate \(\mu_k(z;\theta)\) depends on a rotation parameter \(\theta_r\), the parameter-shift rule gives the identical gradient
\begin{equation}
\frac{\partial \mu_k(z;\theta)}{\partial \theta_r}
=
\frac{1}{2}
\left[
\mu_k\!\left(z;\theta+\frac{\pi}{2}e_r\right)
-
\mu_k\!\left(z;\theta-\frac{\pi}{2}e_r\right)
\right],
\end{equation}
with all other parameters fixed.
This rule is what execution on quantum hardware would use in place of simulator backprop; in either case the circuit gradients are propagated through the classical head by the chain rule and combined with standard backpropagation to optimize the downstream loss \(\mathcal{L}\).
Thus, the measured feature vector \(q_\theta(z)\) and the task head are optimized end-to-end while the LM backbone remains frozen.

\subsection{PennyLane-based SQUARE Circuit Implementation}
\label{app:square_pennylane_code}

Listing~\ref{lst:square_pennylane} shows a compact PennyLane implementation of SQUARE.

\textbf{Normalization and the zero vector.} The mathematical feature map and the PennyLane path use $\alpha(z)=z/\|z\|_2$ for $z\neq0$; the all-zero vector is outside this map and did not occur in the data. The separate native-PyTorch timing implementation used $z/(\|z\|_2+10^{-8})$ for numerical stabilization. It is therefore only an approximation to the normalized state map (feature discrepancy $4.8\times10^{-9}$ in the reported check), and exact normalization identities refer to the mathematical/PennyLane path, not to that timing-only approximation.

The circuit receives a compact bottleneck vector \(z\), amplitude-encodes it into an \(n\)-qubit state, applies the \(RYRZ\)-Ring-\(RY\) parameterized quantum circuit, and returns the measured SQUARE feature vector.
The trainable tensor \texttt{theta} has shape \((L,n,3)\), where \(L\) is the circuit depth and \(n\) is the number of qubits.
For each layer and qubit, the three parameters correspond to the first \(RY\) rotation, the \(RZ\) rotation, and the final \(RY\) rotation after ring entanglement.
The ring block applies CNOT gates in the order \(1\rightarrow2,\ldots,n-1\rightarrow n,n\rightarrow1\), using \(n\) CNOT gates per layer.
After the circuit evolution, SQUARE concatenates computational-basis probabilities and local Pauli-\(Z\) expectations, matching the measured feature map \(q_\theta(z)\) used by the downstream task head. The QNode is declared with \texttt{interface="torch"} and \texttt{diff\_method="backprop"} and returns \texttt{torch} tensors, so the listing is the actual differentiable training path (gradients reach \texttt{theta} through the statevector) rather than an illustrative forward pass.

\begin{lstlisting}[style=pythonstyle, caption={Compact PennyLane implementation of SQUARE.}, label={lst:square_pennylane}]
import pennylane as qml
import torch

def square_layer(theta_layer, wires):
    """One RYRZ-Ring-RY SQUARE layer."""
    n_qubits = len(wires)

    # Local RYRZ rotation block
    for q, w in enumerate(wires):
        qml.RY(theta_layer[q, 0], wires=w)
        qml.RZ(theta_layer[q, 1], wires=w)

    # Ring CNOT entanglement
    for q in range(n_qubits - 1):
        qml.CNOT(wires=[wires[q], wires[q + 1]])
    qml.CNOT(wires=[wires[-1], wires[0]])

    # Final local RY rotation block
    for q, w in enumerate(wires):
        qml.RY(theta_layer[q, 2], wires=w)


def make_square_qnode(n_qubits, depth):
    """Construct a SQUARE QNode with probability and Pauli-Z readout."""
    wires = list(range(n_qubits))
    dev = qml.device("default.qubit", wires=n_qubits)

    # Differentiable training path used in all experiments:
    # exact reverse-mode gradients through the statevector simulation.
    @qml.qnode(dev, interface="torch", diff_method="backprop")
    def square_circuit(z, theta):
        # Amplitude encoding into 2^n-dimensional Hilbert space
        qml.AmplitudeEmbedding(
            features=z,
            wires=wires,
            normalize=True,
            pad_with=0.0,
        )

        # Parameterized quantum circuit
        for ell in range(depth):
            square_layer(theta[ell], wires)

        # Measured SQUARE feature map q_theta(z)
        probs = qml.probs(wires=wires)
        z_exps = [qml.expval(qml.PauliZ(w)) for w in wires]
        return probs, z_exps

    return square_circuit


def square_features(z, theta):
    """Return concatenated SQUARE features [basis probabilities; Z expectations]."""
    depth, n_qubits, _ = theta.shape
    qnode = make_square_qnode(n_qubits=n_qubits, depth=depth)

    probs, z_exps = qnode(z, theta)          # torch tensors; gradient path to theta is preserved
    return torch.cat([probs, torch.stack(z_exps)])
\end{lstlisting}

\subsection{Trends in Recent Quantum-Enhanced Learning Studies}
\label{app:rel}
\begin{table}[h]
\centering
\scriptsize
\caption{Quantum-enhanced adaptation methods for LMs.}
\label{tab:rel}
\setlength{\tabcolsep}{4pt}
\renewcommand{\arraystretch}{1.08}
\begin{tabular}{lccc}
\toprule
\textbf{Characteristic} & \textbf{QAA} & \textbf{QPA} & \textbf{SQUARE (Ours)} \\
\midrule
Purpose & Adapt states & Generate weights & Relation modeling \\
Data encoding & Amplitude & None & Amplitude \\
Hybrid QC & \CheckmarkBold & \CheckmarkBold & \CheckmarkBold \\
Quantum simulator & \CheckmarkBold & \CheckmarkBold & \CheckmarkBold \\
Experiment & NLG & Perplexity eval. & Controlled / Reranking\\
\bottomrule
\end{tabular}
\end{table}

Table~\ref{tab:appendix_qml_settings_check} summarizes recent quantum-enhanced learning studies from major machine learning venues along three experimental dimensions: whether the method uses a hybrid quantum--classical architecture, how classical data are encoded into quantum states or circuits, and whether evaluation is conducted on a simulator. The table shows that recent QML systems are predominantly hybrid rather than standalone quantum models. Across reinforcement learning, computer vision, graph learning, federated learning, and general ML tasks, quantum circuits are typically used as compact feature processors, trainable transformation modules, or structured components inside larger classical pipelines.
A second trend is that classical-to-quantum encoding is a central design choice. The surveyed studies use a range of encodings, including amplitude, angle, feature-map, positional, re-uploading, block, and tensor-network-inspired representations. This variation suggests that QML performance depends not only on adding a quantum circuit, but also on how classical information is embedded before circuit evolution and measurement. SQUARE follows this design pattern by amplitude-encoding compact frozen LM bottleneck features and using the PQC readout to expose nonlinear interaction features.
A third trend is the continued reliance on simulator-based evaluation. All studies in Table~\ref{tab:appendix_qml_settings_check} explicitly use or discuss quantum simulators, reflecting the current difficulty of evaluating learning pipelines directly on large-scale, fault-tolerant quantum hardware. SQUARE is aligned with this practice: our experiments evaluate the representational utility of measured quantum features in simulation, not near-term hardware advantage. This is also why the paper separately reports simulated noise sensitivity while avoiding claims about real-device robustness.

Finally, Table~\ref{tab:appendix_qml_settings_check} highlights that QML applications have expanded beyond traditional quantum benchmarks into RL, CV, graph learning, FL, and now NLP. SQUARE contributes to this direction by placing a compact PQC at the representation-interaction level of frozen language models. Unlike quantum parameter generation or generic activation adaptation, SQUARE directly operates on frozen semantic bottleneck features and measures relation-sensitive readouts for downstream prediction.
\begin{table*}[h]
\centering
\scriptsize
\caption{
Experimental settings in recent quantum-enhanced learning studies.
A checkmark indicates that the corresponding setting is explicitly used or discussed.
Recent QML studies commonly adopt hybrid quantum--classical architectures, task-specific classical-to-quantum encodings, and simulator-based evaluation.
}
\label{tab:appendix_qml_settings_check}
\setlength{\tabcolsep}{4pt}
\renewcommand{\arraystretch}{1.12}
\resizebox{\textwidth}{!}{
\begin{tabular}{lccccc}
\toprule
\textbf{Work}
& \textbf{Conf}
& \textbf{Category}
& \textbf{Hybrid Q-C}
& \textbf{Encoding}
& \textbf{Simulator} \\
\midrule

TensorRL-QAS~\citep{NeurIPS2025TensorRL}
& NeurIPS & RL & \CheckmarkBold & Matrix Product State & \CheckmarkBold \\

QVF~\citep{NeurIPS2025Quantum}
& NeurIPS & CV & \CheckmarkBold & Amplitude & \CheckmarkBold \\

AQC-DRC~\citep{NeurIPS2025adap}
& NeurIPS & ML & \CheckmarkBold & Amplitude & \CheckmarkBold \\

QDSFormer~\citep{NeurIPS2025QDSFormer}
& NeurIPS & CV & \CheckmarkBold & Angle & \CheckmarkBold \\

QuanONet~\citep{ICML2025QuanOnet}
& ICML & ML & \CheckmarkBold & Feature Map & \CheckmarkBold \\

QRL~\citep{ICML2025QRL}
& ICML & RL & \CheckmarkBold & Angle/Re-uploading & \CheckmarkBold \\

QPE-GNN~\citep{ICML2024QPE-GNN}
& ICML & ML & \CheckmarkBold & Positional & \CheckmarkBold \\

Quorus~\citep{ICLR2026Quorus}
& ICLR & FL & \CheckmarkBold & Amplitude & \CheckmarkBold \\

QNN loss landscapes~\citep{ICLR2025QNN}
& ICLR & ML & \CheckmarkBold & Block & \CheckmarkBold \\

eQMARL~\citep{ICLR2025eQMARL}
& ICLR & RL & \CheckmarkBold & Angle & \CheckmarkBold \\

\textbf{SQUARE (Ours)}
& -- & NLP & \CheckmarkBold & Amplitude & \CheckmarkBold \\

\bottomrule
\end{tabular}
}
\end{table*}

\section{Additional Results}

\subsection{Decision-Boundary Evaluation Details}
\label{app:boundary_details}
This appendix provides the per-geometry boundary-specific comparison summarized in main-text Table~\ref{tab:boundary_lowsup} (Table~\ref{tab:boundary_diag}); the per-method Accuracy/F1/AUC breakdown on the two-dimensional PCA boundary task is reported in main-text Table~\ref{tab:adapter_val_results}.

\begin{table}[h]
\centering
\scriptsize
\setlength{\tabcolsep}{4pt}\renewcommand{\arraystretch}{0.95}
\caption{Boundary-specific controlled interaction recovery under an identical validation-selection protocol. Values are mean test AUC $\pm$ sample standard deviation over six random seeds.}
\label{tab:boundary_diag}
\begin{tabular}{l|c|ccc}
\toprule
Method & Params & Checkerboard & Radial ring & Average \\
\midrule
Classical Fourier & 210 & 0.977 $\pm$ 0.033 & 0.983 $\pm$ 0.004 & 0.980 \\
Explicit polynomial & 386 & 0.973 $\pm$ 0.001 & 0.977 $\pm$ 0.003 & 0.975 \\
MLP (GELU) & 961 & 0.487 $\pm$ 0.009 & 0.710 $\pm$ 0.041 & 0.598 \\
Adapter transformation & 140 & 0.505 $\pm$ 0.071 & 0.638 $\pm$ 0.012 & 0.571 \\
LoRA transformation & 77 & 0.567 $\pm$ 0.025 & 0.517 $\pm$ 0.009 & 0.542 \\
\textbf{SQUARE (Ours)} & 281 & \textbf{0.987 $\pm$ 0.009} & \textbf{0.997 $\pm$ 0.002} & \textbf{0.992} \\
\bottomrule
\end{tabular}
\end{table}

\subsection{Compact-Budget and Original-Label GLUE Comparisons}
\label{app:glue_extra}

This section separates two complementary questions. First, under controlled interaction labels, can SQUARE retain competitive accuracy when all methods are restricted to a genuinely compact trainable budget? Second, when the original GLUE labels are restored, how much predictive utility remains when the language model and bottleneck are frozen? The first comparison tests the parameterization of interactions; the second tests whether the resulting adapter remains useful outside the constructed-label setting.

\begin{table}[h]
\centering
\scriptsize
\setlength{\tabcolsep}{3.5pt}
\renewcommand{\arraystretch}{0.95}
\caption{Compact-budget comparison on four GLUE-derived controlled-label tasks. Hyperparameters are selected using validation data only, with a target budget of at most $1.5\times$ the smallest SQUARE configuration. Values are mean $\pm$ s.d.\ over the matched runs. $^{*}$The smallest prespecified configuration evaluated for this method exceeds the target; these rows are retained as informative higher-budget references. In particular, the reported MLP result should not be interpreted as a lower bound on the size of an MLP.}
\label{tab:compact_budget}
\begin{tabular}{l|c|ccccc}
\toprule
Method & Params & SST-2 & RTE & MRPC & QNLI & Avg. \\
\midrule
MLP$^{*}$ & 577
& 0.964 $\pm$ 0.002
& 0.976 $\pm$ 0.002
& 0.958 $\pm$ 0.002
& 0.956 $\pm$ 0.003
& 0.964 \\

Explicit norm.\ quadratic$^{*}$ & 409
& 0.799 $\pm$ 0.006
& 0.859 $\pm$ 0.002
& 0.854 $\pm$ 0.004
& 0.827 $\pm$ 0.005
& 0.835 \\

Full bilinear$^{*}$ & 273
& 0.874 $\pm$ 0.027
& 0.918 $\pm$ 0.019
& 0.902 $\pm$ 0.011
& 0.908 $\pm$ 0.018
& 0.901 \\

Low-rank bilinear & 81
& 0.971 $\pm$ 0.005
& 0.967 $\pm$ 0.005
& 0.942 $\pm$ 0.011
& 0.962 $\pm$ 0.009
& 0.961 \\

Factorization machine & 71
& \textbf{0.973 $\pm$ 0.002}
& 0.972 $\pm$ 0.002
& 0.965 $\pm$ 0.005
& \textbf{0.963 $\pm$ 0.003}
& 0.968 \\

\textbf{SQUARE (Ours)} & \textbf{68}
& 0.967 $\pm$ 0.008
& \textbf{0.980 $\pm$ 0.005}
& \textbf{0.972 $\pm$ 0.002}
& 0.958 $\pm$ 0.009
& \textbf{0.969} \\
\bottomrule
\end{tabular}
\end{table}

Within the target budget, SQUARE attains the highest macro-average ($0.969$) with the fewest trainable parameters ($68$), while the factorization machine is within $0.001$ using $71$ parameters and the low-rank bilinear model reaches $0.961$ using $81$. SQUARE also gives the strongest observed results on RTE and MRPC. Thus, the evidence is not that quadratic interactions require a quantum circuit; rather, SQUARE realizes a competitive second-order inductive bias with a particularly small parameterization. The close factorization-machine result further supports our central interpretation that the structure imposed on interaction coefficients, rather than parameter count alone, is the relevant design choice.

\begin{table}[h]
\centering
\scriptsize
\setlength{\tabcolsep}{4pt}
\renewcommand{\arraystretch}{0.95}
\caption{Performance on the original GLUE labels. SST-2, RTE, and QNLI use accuracy, whereas MRPC uses F1; Avg.\ is their unweighted mean and is not directly comparable to the controlled-label results. Encoder-adapted LoRA updates the backbone, whereas the remaining methods operate on the same frozen bottleneck.}
\label{tab:original_glue}
\begin{tabular}{l|l|c|ccccc}
\toprule
Protocol & Method & Params & SST-2 & RTE & MRPC & QNLI & Avg. \\
\midrule
Encoder-adapted
& LoRA
& 221,953
& 0.870
& 0.581
& 0.832
& 0.736
& 0.755 \\

Frozen bottleneck
& MLP
& 5,041
& 0.822
& 0.540
& 0.811
& 0.571
& 0.686 \\

Frozen bottleneck
& \textbf{SQUARE (Ours)}
& \textbf{49}
& \textbf{0.826}
& 0.527
& \textbf{0.815}
& 0.578
& \textbf{0.686} \\

Frozen bottleneck
& Explicit quadratic
& 445
& 0.824
& 0.475
& 0.813
& 0.580
& 0.673 \\
\bottomrule
\end{tabular}
\end{table}

With the original labels, SQUARE matches the frozen-bottleneck MLP average ($0.686$) using $49$ rather than $5{,}041$ trainable parameters---approximately $103\times$ fewer---and slightly improves SST-2 and MRPC within that frozen-bottleneck comparison. Encoder-adapted LoRA remains more accurate because it updates the language-model representation and uses over $220$K trainable parameters, so it answers a different adaptation question. These results do not establish an original-label accuracy advantage; they show that the compact measured feature map preserves the utility of a much larger frozen-bottleneck head. Together with Table~\ref{tab:compact_budget}, this supports SQUARE as a parameter-efficient interaction adapter rather than as a replacement for full encoder adaptation.

\subsection{Post-Training Simulated Noise Sensitivity}
\label{app:noise_sensitivity}
\vspace{-2mm}

\begin{figure}[h]
\centering
\includegraphics[width=0.7\linewidth]{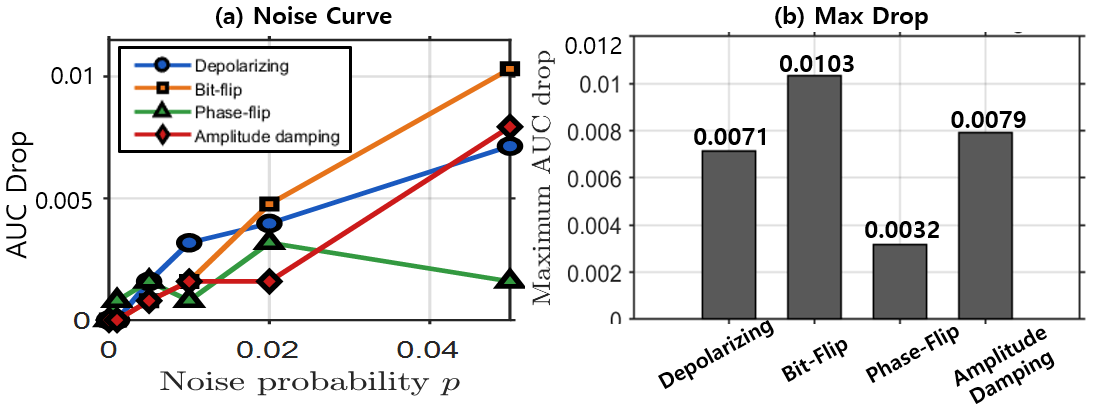}
\caption{
Post-training simulated quantum noise sensitivity of SQUARE with the BERT frozen bottleneck in MNLI.
A trained SQUARE model is evaluated under depolarizing, bit-flip, phase-flip, and amplitude-damping noise injected only during evaluation~\citep{quantum2025amplitude}.
}
\label{fig:noise_sensitivity_main}
\end{figure}

We additionally test whether the learned SQUARE readout remains stable when standard quantum noise channels are introduced after clean training.
This evaluation isolates inference-time noise sensitivity: the model is trained once on a noiseless simulator, its parameters are fixed, and only the quantum circuit evaluation is perturbed.
The purpose is to assess simulator-based sensitivity of the measured feature map, rather than to claim robustness on physical quantum hardware.
We inject depolarizing, bit-flip, phase-flip, and amplitude-damping noise during evaluation and report the resulting AUC drop from the clean model.

Fig.~\ref{fig:noise_sensitivity_main} shows mild degradation across all tested channels.
The largest AUC drop is \(0.0103\) under bit-flip noise, while depolarizing, amplitude-damping, and phase-flip noise yield maximum drops of \(0.0071\), \(0.0079\), and \(0.0032\), respectively.
The non-monotonic trend for some channels reflects the finite validation set and simulator-level stochastic perturbations, so we interpret these results as a sensitivity diagnostic rather than a calibrated hardware-noise benchmark.
SQUARE's measured readout remains reasonably stable under the evaluated post-training noise channels.
%This stability suggests that the measured feature representation is not overly sensitive to small simulator-injected perturbations in this setting, although further hardware-aware evaluation is needed to determine whether the same behavior holds on real quantum devices.

\subsection{Broad Frozen-Bottleneck Approximator Suite}
\label{app:broad_family}
Table~\ref{tab:broad_family} evaluates a broad set of kernel, non-parametric, graph-based, manifold-based, and explicit nonlinear approximators under a shared frozen-bottleneck protocol. All methods use the same data, bottleneck representations, splits, labels, and evaluation metric; each family is selected from prespecified configurations using validation data only. Reported learned-scalar counts are descriptive and are not used to establish a cross-family parameter-efficiency ranking, because non-parametric and kernel methods do not admit a directly comparable count. SQUARE has the highest observed mean AUC, but the uncertainties of several leading methods overlap; we therefore place SQUARE within the strongest observed performance group rather than claiming a statistically established total ordering.

\begin{table}[h]
\centering
\footnotesize
\caption{Broad-family comparison under a shared frozen-bottleneck protocol. Results are mean test AUC $\pm$ sample standard deviation over two interaction rules and six seeds. ``---'' indicates that a directly comparable learned-scalar count is not reported.}
\label{tab:broad_family}
\begin{tabular}{l|c|c}
\toprule
Method & Reported learned scalars & Test AUC \\
\midrule
\textbf{SQUARE (Ours)} & \textbf{234} & \textbf{0.984 $\pm$ 0.014} \\
SVM (RBF) & --- & 0.974 $\pm$ 0.030 \\
Classical Fourier & 204 & 0.961 $\pm$ 0.050 \\
Graph-Laplacian & --- & 0.953 $\pm$ 0.003 \\
Explicit polynomial & 334 & 0.947 $\pm$ 0.030 \\
kNN ($k=25$) & --- & 0.909 $\pm$ 0.011 \\
Random Fourier features & --- & 0.905 $\pm$ 0.103 \\
Nystr\"om & --- & 0.846 $\pm$ 0.146 \\
MLP & 785 & 0.603 $\pm$ 0.136 \\
Factorization machine & 18 & 0.530 $\pm$ 0.015 \\
Adapter transformation & 160 & 0.525 $\pm$ 0.087 \\
Local manifold (Isomap) & --- & 0.517 $\pm$ 0.035 \\
Bilinear & 37 & 0.508 $\pm$ 0.044 \\
\bottomrule
\end{tabular}
\end{table}

\subsection{Additional Backbone Results}
\label{app:exp}
\begin{table}[h]
\centering
\scriptsize
\caption{Validation-selected controlled-interaction accuracy across four frozen LM backbones. Gain denotes SQUARE minus MLP within each backbone and protocol. OPT-350M and GPT-2 use the V20 configuration, whereas OpenLLaMA-3B and Mistral-7B use H16; parameter counts therefore differ across the two groups. These results test whether the qualitative comparison persists across frozen representation geometries and are not directly comparable to the primary disjoint held-out test in Table~\ref{tab:heldout_evidence_summary}.}
\label{tab:backbone_summary}
\setlength{\tabcolsep}{4.2pt}\renewcommand{\arraystretch}{0.96}
\begin{tabular}{lrrrrr}
\toprule
Backbone & MLP & SQUARE & Gain & MLP params & SQUARE params \\
\midrule
OPT-350M     & 0.7960 & \textbf{0.8294} & $+0.0334$ & 776   & 180 \\
GPT-2        & 0.7847 & \textbf{0.8258} & $+0.0411$ & 776   & 180 \\
\midrule
OpenLLaMA-3B & 0.7052 & \textbf{0.7714} & $+0.0662$ & 1,633 & 157 \\
Mistral-7B   & 0.7012 & \textbf{0.7481} & $+0.0469$ & 1,633 & 157 \\
\bottomrule
\end{tabular}
\end{table}

Table~\ref{tab:backbone_summary} consolidates the protocol-level averages before the task-wise breakdowns below. Across all four frozen backbones, SQUARE retains the same qualitative ordering relative to the corresponding MLP while using fewer trainable post-bottleneck parameters. Because labels and adapters are regenerated for each representation, the table is evidence of repeated within-backbone behavior rather than direct transfer between language models.

\begin{figure}[h]
\centering
\includegraphics[width=0.99\linewidth]{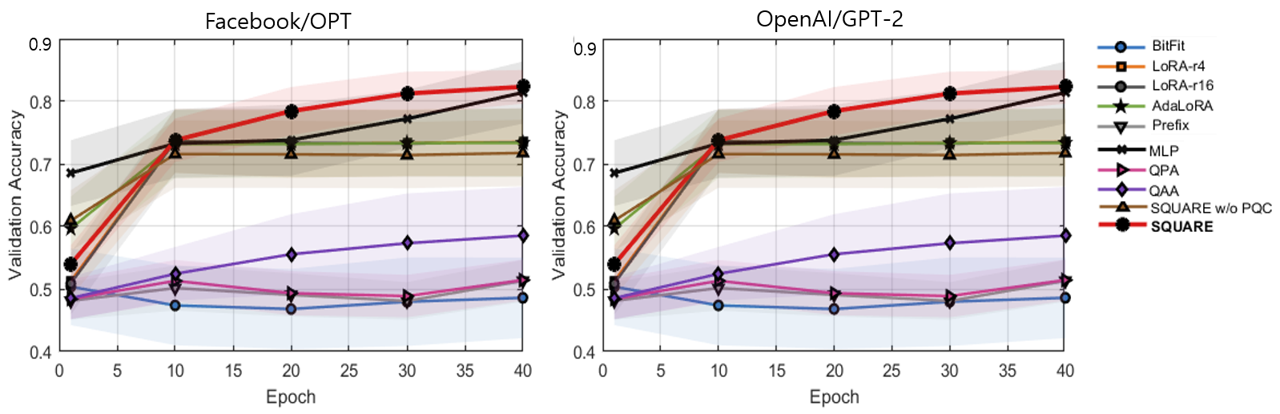}
\caption{
Learning curves on GLUE-derived controlled nonlinear interaction classification with additional frozen backbones.
We report validation accuracy averaged across controlled GLUE-derived tasks for OPT-350M and GPT-2.
Shaded regions denote variation across tasks.
Across both frozen backbones, SQUARE shows stable optimization and reaches the strongest final validation accuracy, while MLP remains the strongest classical baseline.
The gap between SQUARE and SQUARE without the PQC suggests that the measured quantum feature map contributes beyond the lightweight downstream head.
}
\label{fig:additional_backbone_learning_curves}
\end{figure}

\BfPara{Learning Dynamics on Additional Backbones}
Fig.~\ref{fig:additional_backbone_learning_curves} shows the validation learning curves for OPT-350M and GPT-2 on the GLUE-derived controlled interaction tasks.
The results are averaged across tasks, and the shaded regions indicate task-wise variation.
Across both frozen backbones, SQUARE exhibits stable optimization and consistently improves over training.
After the early training phase, SQUARE reaches the top-performing region and obtains the strongest final validation accuracy among the compared adapters.

The trend is consistent across OPT-350M and GPT-2.
MLP remains a strong classical nonlinear baseline, but SQUARE reaches higher final accuracy while using a substantially smaller adapter.
In contrast, QAA, QPA, and SQUARE without the PQC remain clearly below SQUARE.
This suggests that the improvement is not explained merely by adding a quantum-labeled component or a lightweight head; rather, the measured quadratic feature map contributes beyond the lightweight head alone.

These learning curves complement the task-wise results in Tables~\ref{tab:opt350m_glue_taskwise_acc_params} and~\ref{tab:gpt2_glue_taskwise_acc_params}.
They show that SQUARE's gains are not due to unstable late-epoch fluctuations, but arise from a stable training trajectory that appears across different frozen LM geometries.

\begin{table*}[h]
\centering
\footnotesize
\caption{
Task-wise validation accuracy on GLUE-derived controlled nonlinear interaction classification with OPT-350M.
All methods use the same frozen OPT-350M bottleneck representation, and Params counts only trainable adapter parameters.
Several baselines approach majority-class performance under this protocol.
}
\label{tab:opt350m_glue_taskwise_acc_params}
\resizebox{\textwidth}{!}{
\begin{tabular}{l|c|c|rrrrrrrrr}
\toprule
Method & Quantum & Params & CoLA & SST-2 & STS-B & QQP & MNLI & QNLI & RTE & WNLI & Avg. \\
\midrule

BitFit & \XSolidBrush
& 33
& 0.4775 & 0.4825 & 0.4325 & 0.4700 & 0.5125 & 0.4600 & 0.5379 & 0.5070 & 0.4850 \\

LoRA$_{r{=}4}$ & \XSolidBrush
& 104
& 0.6000 & 0.6050 & 0.6450 & 0.6700 & 0.7225 & 0.7675 & 0.8159 & 0.7887 & 0.7018 \\

LoRA$_{r{=}8}$ & \XSolidBrush
& 200
& 0.6500 & 0.6125 & 0.6460 & 0.6700 & 0.7275 & 0.7679 & 0.8231 & 0.7928 & 0.7112 \\

AdaLoRA & \XSolidBrush
& 621
& 0.6850 & 0.6025 & 0.6450 & 0.6675 & 0.7200 & 0.7675 & 0.8303 & 0.7324 & 0.7063 \\

Prefix & \XSolidBrush
& 289
& 0.4775 & 0.4825 & 0.4325 & 0.4700 & 0.5125 & 0.4600 & 0.5379 & 0.5070 & 0.4850 \\

MLP & \XSolidBrush
& 776
& 0.7775 & 0.8025 & 0.7925 & 0.7950 & 0.7825 &  \textbf{0.7775} & 0.8520 & 0.7887 & 0.7960 \\

\midrule

QAA & \CheckmarkBold
& 140
& 0.5500 & 0.6350 & 0.4325 & 0.4900 & 0.5100 & 0.5150 & 0.5379 & 0.4930 & 0.5204 \\

QPA & \CheckmarkBold
& 192
& 0.4775 & 0.4825 & 0.4325 & 0.4700 & 0.5125 & 0.4600 & 0.5379 & 0.5070 & 0.4850 \\

SQUARE (w/o PQC) & \CheckmarkBold
& 168
& 0.6800 & 0.5875 & 0.5475 & 0.6725 & 0.7250 & 0.7650 & 0.8267 & 0.7606 & 0.6956 \\

\textbf{SQUARE (Ours)} & \CheckmarkBold
& 180
& \textbf{0.8075} & \textbf{0.8825} & \textbf{0.8375} & \textbf{0.8500} & \textbf{0.8400} & 0.7700 & \textbf{0.8556} & \textbf{0.7924} & \textbf{0.8294} \\
\bottomrule
\end{tabular}
}
\end{table*}

To evaluate whether SQUARE depends on a specific frozen encoder, we repeat the GLUE-derived controlled nonlinear interaction classification experiment with additional frozen LM backbones, OPT-350M and GPT-2. The experimental protocol is identical to the main BERT-based setting: original GLUE labels are discarded, controlled nonlinear labels are generated over frozen bottleneck representations, and only adapter parameters are trained. This isolates the effect of the adapter under different frozen representation geometries.

Tables~\ref{tab:opt350m_glue_taskwise_acc_params} and~\ref{tab:gpt2_glue_taskwise_acc_params} report task-wise validation accuracy. With OPT-350M, SQUARE achieves the best average accuracy, improving over the strongest classical baseline MLP from \(0.7960\) to \(0.8294\). SQUARE also improves substantially over SQUARE without the PQC, from \(0.6956\) to \(0.8294\), indicating that the measured quantum feature map remains useful beyond the BERT backbone.

\begin{table*}[h]
\centering
\footnotesize
\caption{
Task-wise validation accuracy on GLUE-derived controlled nonlinear interaction classification with GPT-2.
All methods use the same frozen GPT-2 bottleneck representation, and Params counts only trainable adapter parameters.
Several weak baselines collapse to near-majority-class prediction on this backbone and return identical task-wise accuracies; the rows are retained for completeness.
}
\label{tab:gpt2_glue_taskwise_acc_params}
\resizebox{\textwidth}{!}{
\begin{tabular}{l|c|c|rrrrrrrrr}
\toprule
Method & Quantum & Params & CoLA & SST-2 & STS-B & QQP & MNLI & QNLI & RTE & WNLI & Avg. \\
\midrule

BitFit & \XSolidBrush
& 33
& 0.4825 & 0.4900 & 0.5075 & 0.4425 & 0.4375 & 0.4525 & 0.4946 & 0.5070 & 0.4768 \\

LoRA$_{r{=}4}$ & \XSolidBrush
& 104
& 0.7700 & 0.7425 & 0.7800 & 0.6300 & 0.7525 & 0.7325 & 0.6029 & 0.7528 & 0.7204 \\

LoRA$_{r{=}8}$ & \XSolidBrush
& 200
& 0.7775 & 0.7475 & 0.7875 & 0.6675 & 0.7550 & 0.7325 & 0.6282 & 0.7810 & 0.7346 \\

AdaLoRA & \XSolidBrush
& 621
& 0.7750 & 0.7575 & 0.7825 & 0.6375 & 0.7550 & 0.7325 & 0.6534 & 0.8028 & 0.7370 \\

Prefix & \XSolidBrush
& 289
& 0.4850 & 0.4900 & 0.5075 & 0.4425 & 0.4375 & 0.4525 & 0.4946 & 0.5070 & 0.4771 \\

MLP & \XSolidBrush
& 776
& 0.7625 & 0.7925 & 0.8300 & 0.6950 & 0.8275  & 0.7675 & 0.7978 & 0.8046 &0.7847 \\

\midrule

QAA & \CheckmarkBold
& 140
& 0.5875 & 0.6750 & 0.6550 & 0.4975 & 0.5700 & 0.5100 & 0.4946 & 0.5070 & 0.5621 \\

QPA & \CheckmarkBold
& 192
& 0.4850 & 0.4900 & 0.5075 & 0.4425 & 0.4375 & 0.4525 & 0.4946 & 0.5070 & 0.4771 \\

SQUARE (w/o PQC) & \CheckmarkBold
& 168
& 0.7675 & 0.7450 & 0.7900 & 0.6650 & 0.7400 & 0.7300 & 0.6173 & 0.7887 & 0.7304 \\

\textbf{SQUARE (Ours)} & \CheckmarkBold
& 180
& \textbf{0.8250} & \textbf{0.8500}& \textbf{0.8900} & \textbf{0.8150} & \textbf{0.8375}& \textbf{0.7750} & \textbf{0.8051} & \textbf{0.8087} & \textbf{0.8258} \\

\bottomrule
\end{tabular}
}
\end{table*}

With GPT-2, SQUARE again obtains the best average accuracy, improving over MLP from \(0.7847\) to \(0.8258\). The improvement over SQUARE without the PQC is also clear, from \(0.7304\) to \(0.8258\). These results show that the controlled-task performance pattern is not confined to the BERT bottleneck and persists across the evaluated OPT-350M and GPT-2 representations. Together, these results support the use of the measured feature map across the evaluated frozen-LM bottleneck geometries, while not establishing generalization to arbitrary backbones or natural-label tasks.

At the same time, the relative task-wise gains vary across backbones. This is expected because each frozen LM induces a different bottleneck geometry, and the usefulness of nonlinear interaction features depends on which task-relevant relations are preserved in that representation. Overall, the additional backbone results support the view that SQUARE acts as a compact nonlinear relation module over frozen representations, rather than as a backbone-specific adapter.

\BfPara{Scaling to Billion-Parameter Backbones}
We further scale the controlled-interaction study to two substantially larger frozen decoder backbones, OpenLLaMA-3B~\citep{openlm2023openllama} and Mistral-7B~\citep{jiang2023mistral}, under the same experimental protocol: the backbone is kept frozen, controlled nonlinear labels are constructed over the resulting bottleneck representations, and only the post-bottleneck adapter and task head are trained. Because these decoder-only models do not provide a dedicated \texttt{[CLS]} representation, we mean-pool the final hidden states to obtain the bottleneck representation. Tables~\ref{tab:openllama_glue_taskwise_acc_params} and~\ref{tab:mistral_glue_taskwise_acc_params} report the resulting task-wise validation accuracy averaged over three seeds.

Across both larger backbones, SQUARE achieves the highest macro-average accuracy among the reported methods. On OpenLLaMA-3B, SQUARE reaches an average accuracy of (0.7714), compared with (0.7052) for the substantially larger MLP head. On Mistral-7B, SQUARE similarly achieves (0.7481), compared with (0.7012) for MLP. SQUARE also consistently outperforms the evaluated Linear, BitFit, LoRA, and AdaLoRA baselines on the macro-average while using only 157 trainable parameters, substantially fewer than LoRA, AdaLoRA, and MLP. These results indicate that the effectiveness of the structured measured-quadratic feature map is preserved when the frozen representation is produced by substantially larger language models.

The task-wise results further show that the advantage is not driven by a single task. On OpenLLaMA-3B, SQUARE obtains the highest accuracy on seven of the eight evaluated tasks, with MLP outperforming it only on WNLI. On Mistral-7B, SQUARE obtains the highest accuracy on six of the eight tasks, while MLP performs better on STS-B and WNLI. The magnitude of the gain nevertheless varies across tasks and backbones, which is expected because each frozen language model induces a different bottleneck geometry and therefore preserves different forms of task-relevant interaction structure.

Taken together with the OPT-350M and GPT-2 results, these experiments show that SQUARE's controlled-interaction performance is not restricted to a particular frozen language-model family or model scale. Rather, the measured quadratic feature map remains effective across the evaluated frozen representation geometries, ranging from smaller OPT and GPT-2 backbones to billion-parameter OpenLLaMA and Mistral models. Importantly, these results are specific to the controlled nonlinear interaction classification setting and do not establish that SQUARE is universally superior across arbitrary downstream objectives or natural-label tasks.

\begin{table*}[h]
\centering
\footnotesize
\caption{
Task-wise validation accuracy on GLUE-derived controlled nonlinear interaction classification with a frozen \textbf{OpenLLaMA-3B} backbone, averaged over three seeds. All methods operate on the same frozen OpenLLaMA-3B bottleneck representation. Params counts trainable adapter and task-head parameters only and therefore differs from the main-text budget because the head width and adapter ranks are configured for this backbone. Best performance in each column is shown in \textbf{bold}. SQUARE achieves the highest macro-average accuracy while using substantially fewer trainable parameters than LoRA, AdaLoRA, and the larger MLP head.
}
\label{tab:openllama_glue_taskwise_acc_params}
\resizebox{\textwidth}{!}{
\begin{tabular}{l|c|c|rrrrrrrr|r}
\toprule
Method & Quantum & Params & CoLA & SST-2 & STS-B & QQP & MNLI & QNLI & RTE & WNLI & Avg. \\
\midrule
Linear & \XSolidBrush & 17 & 0.6358 & 0.5783 & 0.6083 & 0.6583 & 0.6083 & 0.6392 & 0.6474 & 0.6150 & 0.6238 \\
BitFit & \XSolidBrush & 33 & 0.4808 & 0.5375 & 0.5592 & 0.5158 & 0.4717 & 0.4908 & 0.5535 & 0.5352 & 0.5181 \\
LoRA$_{r{=}8}$ & \XSolidBrush & 417 & 0.6300 & 0.5825 & 0.6058 & 0.6525 & 0.6075 & 0.6417 & 0.6450 & 0.6103 & 0.6219 \\
AdaLoRA & \XSolidBrush & 621 & 0.6292 & 0.5725 & 0.5892 & 0.6592 & 0.6050 & 0.6400 & 0.6510 & 0.6103 & 0.6195 \\
% Prefix & \XSolidBrush & 289 & 0.7300 & 0.7325 & \textbf{0.8133} & 0.7800 & 0.7600 & 0.7600 & 0.8111 & 0.7747 & 0.7702 \\
MLP & \XSolidBrush & 1633 & 0.6667 & 0.6425 & 0.7442 & 0.6858 & 0.6767 & 0.6367 & 0.6967 & \textbf{0.8920} & 0.7052 \\
\midrule
\textbf{SQUARE (Ours)} & \CheckmarkBold & 157 & \textbf{0.7667} & \textbf{0.7358} & \textbf{0.7633} & \textbf{0.7825} & \textbf{0.7833} & \textbf{0.8025} & \textbf{0.8231} & 0.7136 & \textbf{0.7714} \\
\bottomrule
\end{tabular}
}
\end{table*}

\begin{table*}[h]
\centering
\footnotesize
\caption{
Task-wise validation accuracy on GLUE-derived controlled nonlinear interaction classification with a frozen \textbf{Mistral-7B-v0.1} backbone, averaged over three seeds. All methods operate on the same frozen Mistral-7B-v0.1 bottleneck representation. Params counts trainable adapter and task-head parameters only and therefore differs from the main-text budget because the head width and adapter ranks are configured for this backbone. Best performance in each column is shown in \textbf{bold}. SQUARE achieves the highest macro-average accuracy while using substantially fewer trainable parameters than LoRA, AdaLoRA, and the larger MLP head.
}
\label{tab:mistral_glue_taskwise_acc_params}
\resizebox{\textwidth}{!}{
\begin{tabular}{l|c|c|rrrrrrrr|r}
\toprule
Method & Quantum & Params & CoLA & SST-2 & STS-B & QQP & MNLI & QNLI & RTE & WNLI & Avg. \\
\midrule
Linear & \XSolidBrush & 17 & 0.6017 & 0.5692 & 0.6167 & 0.6225 & 0.6083 & 0.5758 & 0.5860 & 0.6244 & 0.6006 \\
BitFit & \XSolidBrush & 33 & 0.4950 & 0.4883 & 0.5200 & 0.5517 & 0.4842 & 0.4500 & 0.5090 & 0.5634 & 0.5077 \\
LoRA$_{r{=}8}$ & \XSolidBrush & 417 & 0.5975 & 0.5675 & 0.5800 & 0.6308 & 0.6133 & 0.5783 & 0.5752 & 0.6291 & 0.5965 \\
AdaLoRA & \XSolidBrush & 621 & 0.5958 & 0.5700 & 0.5808 & 0.6233 & 0.6000 & 0.6000 & 0.5909 & 0.5962 & 0.5946 \\
% Prefix & \XSolidBrush & 289 & \textbf{0.7667} & \textbf{0.7550} & \textbf{0.7925} & \textbf{0.7775} & 0.7842 & 0.6983 & \textbf{0.7449} & 0.8498 & \textbf{0.7711} \\
MLP & \XSolidBrush & 1633 & 0.6225 & 0.6592 & 0.7625 & 0.6650 & 0.6567 & 0.6475 & 0.6811 & \textbf{0.9155} & 0.7012 \\
\midrule
\textbf{SQUARE (Ours)} & \CheckmarkBold & 157 & \textbf{0.7442} & \textbf{0.7442} & \textbf{0.7100} & \textbf{0.7158} & \textbf{0.7850} & \textbf{0.7642} & \textbf{0.7377} & 0.7840 & \textbf{0.7481} \\
\bottomrule
\end{tabular}
}
\end{table*}

\begin{table}[h]
\centering
\footnotesize
\caption{Controlled candidate reranking on the two larger frozen backbones using an MRPC-derived controlled reranking task (mean over three seeds). Label Acc./F1 measure the correctness of the selected candidate, and Cand.\ AUC measures candidate-relevance ranking quality. Under a parameter-matched comparison, both SQUARE variants improve candidate-level ranking AUC over the classical MLP on both backbones. SQUARE (Hybrid) further achieves the highest Label Accuracy and F1 on both OpenLLaMA-3B and Mistral-7B-v0.1, indicating that combining the structured quantum representation with the hybrid prediction pathway provides the most consistent reranking performance across the evaluated metrics.}
\label{tab:backbone_reranking}
\begin{tabular}{ll|ccc}
\toprule
Backbone & Reranker & Label Acc. & Label F1 & Cand. AUC \\
\midrule
\multirow{5}{*}{OpenLLaMA-3B} & Random & 0.485 & 0.490 & -- \\
 & Linear & 0.515 & 0.679 & 0.522 \\
 & MLP & 0.707 & 0.701 & 0.736 \\
 & \textbf{SQUARE (Quantum)} & 0.704 & 0.712 & \textbf{0.745} \\
 & SQUARE (Hybrid) & \textbf{0.708} & \textbf{0.724} & \textbf{0.745} \\
\midrule
\multirow{5}{*}{Mistral-7B-v0.1} & Random & 0.510 & 0.491 & -- \\
 & Linear & 0.531 & 0.000 & 0.538 \\
 & MLP & 0.727 & 0.704 & 0.760 \\
 & \textbf{SQUARE (Quantum)} & 0.706 & 0.686 & 0.778 \\
 & SQUARE (Hybrid) & \textbf{0.731 }& \textbf{0.713 }& \textbf{0.779}\\
\bottomrule
\end{tabular}
\end{table}

\BfPara{Controlled Reranking on Larger Backbones}
Table~\ref{tab:backbone_reranking} extends the controlled candidate-reranking analysis to OpenLLaMA-3B and Mistral-7B-v0.1 under a parameter-matched comparison. Across both backbones, the SQUARE-based rerankers consistently improve candidate-level ranking AUC over the classical MLP baseline. On OpenLLaMA-3B, SQUARE (Quantum) and SQUARE (Hybrid) both achieve an AUC of (0.745), compared with (0.736) for MLP. The hybrid variant additionally obtains the highest Label Accuracy and F1, reaching (0.708) and (0.724), respectively, compared with (0.707) and (0.701) for MLP.

A similar pattern is observed on Mistral-7B-v0.1. SQUARE (Hybrid) achieves the strongest performance across all three metrics, with a Label Accuracy of (0.731), an F1 score of (0.713), and a candidate AUC of (0.779), compared with (0.727), (0.704), and (0.760) for MLP. The quantum-only variant also improves candidate-level AUC to (0.778), although its Label Accuracy and F1 remain below those of the MLP and hybrid variants. This distinction suggests that the structured quantum representation provides useful candidate-ranking information, while combining it with the hybrid prediction pathway yields more robust performance for the final label-selection objective.

Together with the controlled nonlinear interaction classification results, these findings show that SQUARE's structured interaction modeling extends beyond direct classification to a candidate-ranking objective. In particular, the consistent AUC gains across both SQUARE variants and both billion-parameter backbones indicate that the measured quantum feature map captures interaction structure useful for relative candidate scoring. At the same time, the stronger Label Accuracy and F1 of the hybrid variant suggest that retaining a complementary classical pathway can improve the conversion of these interaction features into final decisions. We therefore interpret the reranking results as evidence that the proposed structured representation transfers across controlled downstream objectives, while the optimal use of the quantum features may depend on how they are integrated into the prediction head.

\subsection{Compute-Resource Benchmark}
\label{app:compute_resource}

Table~\ref{tab:runtime_compact} reports an adapter-level runtime benchmark with batch size \(32\).
The benchmark measures forward latency, training latency, and training throughput per sample while keeping the frozen backbone representation fixed.
Thus, the reported values reflect the cost of the adaptation module rather than full end-to-end LM training.

\begin{table}[h]
\centering
\footnotesize
\caption{
Adapter-level runtime benchmark with batch size \(32\).
Latency is measured per sample after freezing the backbone representation.
}
\label{tab:runtime_compact}
\setlength{\tabcolsep}{4pt}
\renewcommand{\arraystretch}{1.05}
\begin{tabular}{lrrrr}
\toprule
Method & Params & Fwd. ms & Train ms & Train samples/s \\
\midrule
BitFit & 33 & 0.0036 & 0.0403 & 24812.7 \\
LoRA-\(r{=}4\) & 104 & 0.0060 & 0.0467 & 21430.0 \\
LoRA-\(r{=}8\) & 200 & 0.0059 & 0.0462 & 21627.3 \\
Prefix & 289 & 0.0112 & 0.0660 & 15153.3 \\
MLP & 776 & 0.0058 & 0.0513 & 19477.8 \\
\midrule
QAA & 140 & 8.9023 & 14.7052 & 68.0 \\
QPA & 192 & 0.2606 & 0.5599 & 1786.1 \\
SQUARE & 180 & 11.1924 & 21.3373 & 46.9 \\
\bottomrule
\end{tabular}
\end{table}

Classical adapters are evaluated as standard tensor operations and therefore have very small latency.
For example, BitFit, LoRA, Prefix, and MLP all remain below \(0.07\) ms per sample during training in this benchmark.
In contrast, quantum-enhanced methods require simulator-based circuit evaluation, which introduces substantial overhead.
SQUARE uses only \(180\) trainable parameters, comparable to compact PEFT baselines, but its current PennyLane simulator implementation requires \(21.3373\) ms per sample during training and reaches \(46.9\) training samples/s.

These results highlight a practical trade-off.
SQUARE provides a compact measured nonlinear feature map with a small trainable parameter budget. To make the learned module usable outside the simulator, the direct PyTorch realization preserves the same inputs, parameters, gate order, topology, observables, and batch size as the PennyLane implementation: amplitude encoding becomes normalization and zero-padding; each $RY$/$RZ$ rotation and CNOT is applied as a batched complex-valued matrix--vector operation; basis probabilities are squared magnitudes; and Pauli-$Z$ expectations are signed sums of the probability block. It reproduces the simulated features with maximum absolute difference $1.118\times10^{-7}$ and reduces forward latency from $97.598$~ms to $4.720$~ms at batch size 8 on an NVIDIA H200 ($20.7\times$). This establishes exact, simulator-free deployment of the learned circuit map; it does not imply a speed advantage over specialized classical adapters.

\BfPara{Matched-Protocol Training Benchmark}
Table~\ref{tab:runtime_compact} times per-sample simulator training, which is the configuration the pipeline actually used but not the relevant comparison for the classical realization. We therefore benchmarked \emph{training} (forward, backward, and an Adam step on a binary cross-entropy loss) under one matched protocol: the same $d{=}16$, $n{=}4$ circuit and the same $\mathrm{Lin}(16{\to}8)$--$\tanh$--$\mathrm{Lin}(8{\to}1)$ head, batch size $64$, float64, a single CPU process on the experiment host (PennyLane~0.40.0, PyTorch~2.4.1; \texttt{default.qubit} is CPU-only in our environment; the host was shared with other jobs, so ratios are more informative than absolute times), $10$ warm-up and $100$ timed steps, for three implementations: the PennyLane \texttt{backprop} path, the native PyTorch statevector path (one $16{\times}16$ unitary assembled from the angles per step and applied to the batch), and a $16{\to}8{\to}1$ MLP. On a batch of $64$ random inputs the two SQUARE paths agree to $4.8\times10^{-9}$ in the features and $2.1\times10^{-8}$ in the angle gradients (the residual is the $\varepsilon{=}10^{-8}$ normalization constant). The native path removes most of the simulator overhead but remains slower than the MLP, consistent with the statement that the parameter budget does not establish computational efficiency.
\begin{table}[h]
\centering\footnotesize
\caption{Matched-protocol training cost (CPU, batch $64$, float64, mean $\pm$ s.d.\ over $100$ steps).}
\label{tab:runtime_matched}
\resizebox{\linewidth}{!}{\begin{tabular}{lccc}
\toprule
Implementation & ms / step & ms / sample & relative to MLP \\
\midrule
SQUARE, PennyLane \texttt{default.qubit} + \texttt{backprop} & $14.26\pm1.72$ & $0.223$ & $22.4\times$ \\
SQUARE, native PyTorch statevector & $3.60\pm0.11$ & $0.056$ & $5.7\times$ \\
MLP $16{\to}8{\to}1$ ($145$ parameters) & $0.64\pm0.03$ & $0.010$ & $1.0\times$ \\
\bottomrule
\end{tabular}}
\end{table}

\subsection{Generation-oriented Reranking: Full Results}\label{app:reranking}

\begin{table*}[h]
\centering
\footnotesize
\caption{
Alpaca candidate reranking results.
All methods score the same fixed candidate pool using the same frozen BERT bottleneck; higher is better.
}
\label{tab:alpaca_square_improved}
\setlength{\tabcolsep}{3pt}\renewcommand{\arraystretch}{0.92}
\resizebox{\textwidth}{!}{
\begin{tabular}{l|ccccccc}
\toprule
Selector & BLEU & ROUGE-1 & ROUGE-2 & ROUGE-L & Score & HitRate & NonlinearScore \\
\midrule
Uniform random (exact) & 0.0193 & 0.2078 & 0.0622 & 0.1749 & 0.2066 & 0.1656 & 0.4991 \\
MLP & 0.0215 & 0.2256 & 0.0664 & 0.1861 & 0.2149 & 0.2083 & 0.4917 \\
SQUARE, pure path & 0.0211 & 0.2172 & 0.0605 & 0.1775 & 0.2128 & 0.1917 & 0.5222 \\
\textbf{SQUARE, residual path} & \textbf{0.0246} & \textbf{0.2365} & \textbf{0.0757} & \textbf{0.1895} & \textbf{0.2224} & \textbf{0.2417} & \textbf{0.5032} \\
\midrule
Oracle-$q$ & 0.0288 & 0.3123 & 0.1098 & 0.2622 & 0.2861 & 1.0000 & 0.5662 \\
Oracle-$\rho$ & 0.0172 & 0.1935 & 0.0604 & 0.1618 & 0.2261 & 0.3083 & 0.6890 \\
\bottomrule
\end{tabular}
}
\end{table*}

We formulate Alpaca instruction following as a candidate reranking problem~\citep{CRFM2023Alpaca}: for each instruction, a fixed pool of LLM-generated responses is given, and each method learns to assign a scalar score to each instruction--response pair over the same frozen BERT bottleneck. Because the candidate pool is shared across methods, the evaluation isolates the quality of the learned scorer rather than the generator~\citep{NAACL2024llm, ICML2025Gram}. We evaluate the top-ranked response selected by each method: BLEU and ROUGE measure lexical agreement with the reference, Score and HitRate evaluate whether the scorer selects high-quality candidates, and NonlinearScore measures joint satisfaction of interaction-dependent quality criteria; detailed definitions are provided in Appendix~\ref{app:metrics}.

Table~\ref{tab:alpaca_square_improved} rebuilds the evaluation from candidate-level data for $120$ instructions and $K{=}8$ candidates per pool. Because $11$ pools contain tied maxima, exact random HitRate is $0.1656$ rather than $1/8$. The residual-path SQUARE scorer is numerically highest among the learned scorers, but its differences from the MLP are not separable under instruction-paired bootstrap: $+0.012$ NonlinearScore (95\% CI $[-0.012,0.036]$) and $+0.033$ HitRate ($[-0.033,0.100]$). Random selection already reaches $0.4991$ NonlinearScore, whereas Oracle-$\rho$ reaches $0.6890$; the pure path moves further toward this interaction target ($0.5222$) at the cost of lexical metrics. Accordingly, the experiment supports transfer to a reranking objective and approximate parity with compact classical scoring, but not improved human-perceived quality or a specifically quantum benefit.

\subsection{Limitations and Future Directions}
\label{app:limitations}
Our evaluation primarily emphasizes controlled interaction recovery, which
allows the contribution of the post-bottleneck parameterization to be isolated
from changes in the frozen representation. The original-label and reranking
experiments provide complementary scope checks, while extending the observed
advantages to broader natural-supervision settings remains an important
direction for future work. The eight GLUE-derived entries represent eight frozen-representation distributions under one shared constructed rule rather than eight independent semantic tasks, and the natural-label results currently support compactness rather than an accuracy gain.

Circuit simulation introduces training overhead relative to specialized
classical adapters. SQUARE addresses deployment rather than simulation cost:
after training, the learned map can be compiled exactly into a numerically
equivalent batched PyTorch module that requires neither a circuit simulator nor
quantum hardware at inference. Accordingly, parameter efficiency should not be read as computational efficiency; the native PyTorch realization remains slower than the matched MLP in the reported training benchmark.

Finally, the central experiments use $d{=}16$ bottlenecks and four-qubit statevectors. Although the stored circuit-angle count grows with $O(L\log d)$ for the chosen gate template and the probability readout has width $B<2d$, a dense classical realization of the learned unitary can require $O(B^2)$ storage and arithmetic; few trainable angles therefore do not establish favorable scaling. The additional-backbone study changes the frozen encoder while retaining a compact bottleneck and is not a dimension-scaling experiment. Future work should study bottleneck dimension, alternative circuit and classical mixing structures, finite-shot training, and hardware-aware or gatewise implementations. The current results therefore
support circuit-induced interaction parameterization and exact classical
deployment, without assuming quantum computational or hardware advantage.

\subsection{Broader Impact}
The main intended impact is methodological and practical: SQUARE provides a controlled way to learn interaction-dependent structure through a quantum-circuit parameterization and then deploy the learned map as an exact classical PyTorch module without updating the frozen backbone.
Potential applications include parameter-efficient classification, scoring, and reranking.
The results should not be interpreted as near-term quantum hardware advantage; rather, they demonstrate a train-with-circuit, deploy-classically workflow that makes circuit-induced feature maps accessible to conventional software pipelines.

SQUARE can inherit and potentially amplify biases or spurious correlations in the frozen LM representation.
This risk is especially relevant in sensitive domains such as education, healthcare, hiring, or legal support.
Any deployment should therefore include task-specific validation, fairness and robustness checks, privacy safeguards, and human oversight.
Future work should evaluate SQUARE under natural supervision, larger backbones, stronger privacy constraints, and realistic deployment protocols.

\subsection{Additional Boundary Controls: Untrained-Circuit, Finite-Shot, and Classical Collapse}
\label{app:boundary_controls}
We further evaluate three properties of the boundary pipeline: the contribution
of trained circuit angles, sensitivity to finite-shot readout, and comparison
with mechanism-matched classical controls. The depth-1 SQUARE variant uses a learned nonlinear $2{\to}16$ projection and magnitude features $\sqrt{p}$; as clarified in Appendix~\ref{app:structure_invariance}, this end-to-end boundary architecture is not exactly quadratic in the original PCA-2 coordinates. We compare it with mechanism-matched classical and PEFT-inspired baselines, an RBF-SVM, a frozen-circuit control, and a depth-2 re-uploading variant included as an extended-capacity reference.

\textbf{Checkerboard geometry.} On checkerboard, the MLP, adapter, LoRA, and random forest reach mean test AUC $0.484\pm0.046$, $0.490\pm0.040$, $0.501\pm0.052$, and $0.457\pm0.023$; RBF-SVM reaches $0.609\pm0.013$, and depth-1 SQUARE reaches $0.849\pm0.039$. This is evidence of performance under the stated optimization protocol, not representational impossibility for the baselines and not an isolated test of a quadratic map: the SQUARE boundary pipeline also contains the nonlinear projection and magnitude readout described above. Training-AUC diagnostics would be needed to distinguish failure to fit from failure to generalize.

\textbf{Untrained-circuit control.} In the depth-1 boundary pipeline, trained-minus-untrained AUC gaps are within $[-0.017,+0.022]$ across all five geometries, so trained angles are not decisive under this residual, learned-projection protocol. This observation does not isolate a degree-two end-to-end predictor because the pipeline also contains a nonlinear projection and magnitude readout. On the disjoint held-out pure-path GLUE-derived suite, by contrast, the trained circuit reaches $0.7565$ mean accuracy versus $0.6155$ for the frozen circuit (Appendix~\ref{app:heldout_validation}), a substantially larger separation that isolates the value of learned circuit mixing in the measured-feature path.

\textbf{Finite-shot readout.} Replacing exact-statevector probabilities with multinomial sampling at $100/1{,}000/10{,}000$ shots changes test AUC only slightly: on checkerboard, depth-1 SQUARE moves from $0.849$ to $0.833$ at $100$ shots and returns to $0.849$ at $1{,}000$ and $10{,}000$ shots; the re-uploading variant moves from $0.962$ to $0.954/0.962/0.963$, and other geometries change by at most $\sim0.01$ at $100$ shots. This is a finite-sampling sensitivity check for the stated boundary pipeline, not evidence that its end-to-end predictor is exactly quadratic.

We report these as controls on a single tutoring bottleneck; they complement rather than replace the main-text diagnostics. The depth-1 SQUARE configuration is competitive but not uniformly dominant (on the radial-ring geometry the RBF-SVM and the re-uploading variant are stronger), consistent with our scoping of the contribution as a structured parameterization rather than a universal approximator.

\begin{table}[h]
\centering
\footnotesize
\setlength{\tabcolsep}{3.2pt}\renewcommand{\arraystretch}{0.95}
\caption{Additional boundary controls on a frozen \texttt{Qwen2.5-1.5B-Instruct} tutoring PCA-2 bottleneck. Mean test AUC $\pm$ sample standard deviation over six seeds. SQUARE is the depth-1 learned-projection/magnitude-readout configuration; ``untr.'' freezes all circuit parameters. SQUARE$_{\text{ru}}$ adds depth-2 input re-uploading. These end-to-end boundary models are not claimed to be exactly quadratic in the original PCA-2 coordinates.}
\label{tab:boundary_controls}
\begin{tabular}{l|ccccc}
\toprule
Method & cluster & radial ring & checkerboard & two-moons & islands \\
\midrule
SQUARE (depth-1, $\sim$93p) 
& \textbf{0.999$\pm$0.001} 
& 0.825$\pm$0.045 
& 0.849$\pm$0.039 
& 0.964$\pm$0.034 
& 0.986$\pm$0.013 \\

SQUARE untr. ($\sim$81p) 
& \textbf{0.999$\pm$0.001} 
& 0.841$\pm$0.034 
& 0.838$\pm$0.042 
& 0.981$\pm$0.016 
& 0.964$\pm$0.023 \\

SQUARE$_{\text{ru}}$ (depth-2, $\sim$121p) 
& \textbf{0.999$\pm$0.001} 
& \textbf{0.999$\pm$0.001} 
& \textbf{0.962$\pm$0.033} 
& \textbf{0.999$\pm$0.001} 
& 0.994$\pm$0.004 \\

SQUARE$_{\text{ru}}$ untr. 
& \textbf{0.999$\pm$0.002} 
& 0.933$\pm$0.043 
& 0.829$\pm$0.015 
& 0.997$\pm$0.003 
& 0.988$\pm$0.007 \\
\midrule
MLP ($\sim$129p) 
& \textbf{0.999$\pm$0.001} 
& 0.804$\pm$0.106 
& 0.484$\pm$0.046 
& 0.920$\pm$0.004 
& 0.992$\pm$0.009 \\

Adapter ($\sim$125p) 
& \textbf{0.999$\pm$0.002} 
& 0.773$\pm$0.110 
& 0.490$\pm$0.040 
& 0.920$\pm$0.003 
& 0.994$\pm$0.007 \\

LoRA ($\sim$99p) 
& 0.563$\pm$0.011 
& 0.593$\pm$0.011 
& 0.501$\pm$0.052 
& 0.924$\pm$0.004 
& 0.679$\pm$0.028 \\

RBF-SVM 
& \textbf{0.999$\pm$0.003} 
& 0.996$\pm$0.001 
& 0.609$\pm$0.013 
& 0.966$\pm$0.004 
& \textbf{0.999$\pm$0.002} \\

Random Forest 
& 0.975$\pm$0.014 
& 0.848$\pm$0.032 
& 0.457$\pm$0.023 
& 0.899$\pm$0.023 
& 0.892$\pm$0.009 \\
\bottomrule
\end{tabular}
\end{table}

\begin{table}[h]
\centering
\footnotesize
\setlength{\tabcolsep}{4pt}\renewcommand{\arraystretch}{0.95}
\caption{Finite-shot readout on the tutoring bottleneck: mean test AUC over six seeds when the exact statevector probabilities are replaced by multinomial sampling at $100/1{,}000/10{,}000$ shots. AUC differs from the exact readout by at most $0.016$ at $100$ shots and by $\leq 0.001$ at $\geq 1{,}000$ shots.}
\label{tab:finite_shot}
\begin{tabular}{l|c|ccc}
\toprule
Method (rule) & exact & 100 shots & 1{,}000 & 10{,}000 \\
\midrule
SQUARE (checkerboard) & 0.849 & 0.833 & 0.849 & 0.849 \\
SQUARE$_{\text{ru}}$ (checkerboard) & 0.962 & 0.954 & 0.962 & 0.963 \\
SQUARE$_{\text{ru}}$ (radial ring) & 0.999 & 0.997 & 0.999 & 0.999 \\
SQUARE$_{\text{ru}}$ (two-moons) & 0.999 & 0.999 & 0.999 & 0.999 \\
\bottomrule
\end{tabular}
\end{table}

\subsection{Trainable-Parameter Accounting and Head Architectures for SQUARE}
\label{app:param_accounting}
Several tables report SQUARE with different trainable-parameter totals. All follow $N_{\mathrm{SQUARE}}=gL\lceil\log_2 d\rceil+N_{\mathrm{proj}}+N_{\mathrm{head}}$, where $g$ is the number of trainable rotations per qubit per layer ($g{=}3$ for $RYRZ$--Ring--$RY$, $g{=}1$ or $2$ in the rotation ablation), $N_{\mathrm{proj}}$ counts an optional input projection, and $N_{\mathrm{head}}$ is the task head applied to the readout. Table~\ref{tab:param_accounting} summarizes the head architectures and parameter decompositions of the evaluated SQUARE configurations. The head is a small \emph{nonlinear} ($\tanh$) MLP in the controlled-GLUE suites and affine in the boundary controls, which is why normalized quadraticity applies to the feature map and not to the end-to-end predictor. The validation/backbone suites read all $B{+}n{=}20$ coordinates into a bias-free head, giving $12+160+8=180$, whereas the primary held-out control suite reads the 16 probabilities into a biased head, giving $12+145=157$. Other reported configurations follow the same formula with the head widths given in their table captions and Appendix~\ref{app:baseline_implementation}.

\begin{table}[h]
\centering\footnotesize
\caption{Parameter decomposition of the central SQUARE configurations ($d{=}B{=}16$, $n{=}4$ unless noted). ``Readout'' is the head input dimension. The primary held-out model uses a biased probability-only head; the separate validation/backbone model uses a bias-free probability-plus-$Z$ head. $^{\dagger}$Learned input projection $\mathrm{Lin}(2{\to}16)$--$\tanh$, used only in the boundary suite.}
\label{tab:param_accounting}
\setlength{\tabcolsep}{3pt}\renewcommand{\arraystretch}{1.1}
\begin{tabular}{@{}p{3.7cm}p{4.3cm}rrrr@{}}
\toprule
Protocol & Head architecture & Angles & Proj. & Head & Total \\
\midrule
\raggedright V20: validation/backbone suites and runtime benchmark &
\raggedright $\mathrm{Lin}(20{\to}8)$--$\tanh$--$\mathrm{Lin}(8{\to}1)$, bias-free, on the 20-dimensional readout &
12 & -- & 168 & \textbf{180} \\
\addlinespace
\raggedright H16: primary held-out test and post-entanglement ablation (Apps.~\ref{app:heldout_validation}, \ref{app:corrected_ablation}) &
\raggedright $\mathrm{Lin}(16{\to}8)$--$\tanh$--$\mathrm{Lin}(8{\to}1)$ on the 16 basis probabilities & 12 & -- & 145 & \textbf{157} \\
\addlinespace
\raggedright B32: boundary controls (App.~\ref{app:boundary_controls}), nonlinear projection + magnitude readout &
\raggedright $\mathrm{Lin}(32{\to}1)$ on $[v;\sqrt{p(v)}]$ & 12 & 48$^{\dagger}$ & 33 & \textbf{93} \\
\bottomrule
\end{tabular}
\end{table}

The $20$-dimensional V20 readout contains $16$ probabilities and four Pauli-$Z$ expectations, but the latter are fixed linear combinations of the probabilities. After training, the first-layer weights can therefore be folded exactly into a $16{\to}8$ matrix. The resulting function has an equivalent $148$-scalar inference representation ($12$ angles, $128$ folded first-layer weights, and $8$ output weights), although $180$ remains the correct number of independently optimized scalars in that training configuration. This algebraic compression preserves the trained function. It does not imply identical optimization: training on $[p;\zeta]$ reparameterizes the first-layer gradient relative to training on $p$ alone. We therefore report $180$ as the trained V20 model size and $148$ only as an exact post-training inference representation; H16 is the $157$-parameter primary held-out configuration.

\subsection{Disjoint Held-Out Validation of the Same-Pipeline Controls}
\label{app:heldout_validation}
We evaluate the same-pipeline control suite using disjoint train, validation,
and test subsets on all eight GLUE-derived controlled-label tasks. The scaler, PCA projection, and training-set median label threshold were fitted on train only; the checkpoint was selected by validation AUC and evaluated once on test. Table~\ref{tab:heldout_summary} reports the unweighted eight-task mean over five shared seeds. The $157$-parameter SQUARE total is $12$ circuit angles plus the $145$-parameter biased $16{\to}8{\to}1$ head, as detailed in Table~\ref{tab:param_accounting}.

\begin{table}[h]
\centering\footnotesize
\caption{Disjoint held-out test summary for same-pipeline controls (eight tasks, five shared seeds). $\Delta$ is comparator minus SQUARE. Displayed endpoints are descriptive averages of taskwise example-paired bootstrap 95\% intervals, not a calibrated confidence interval for the macro-average.}
\label{tab:heldout_summary}
\begin{tabular}{lrrc}
\toprule
Method & Params & Avg. test & $\Delta$ vs. SQUARE [avg.\ taskwise interval] \\
\midrule
LoRA $r{=}8$ & 417 & 0.5897 & $-0.167$ [$-0.181,-0.137$] \\
MLP & 1,633 & 0.6817 & $-0.075$ [$-0.081,-0.053$] \\
SQUARE w/o PQC & 145 & 0.6412 & $-0.115$ [$-0.125,-0.092$] \\
SQUARE, frozen circuit & 145 & 0.6155 & $-0.141$ [$-0.147,-0.120$] \\
\textbf{SQUARE} & \textbf{157} & \textbf{0.7565} & -- \\
Givens-12 + squares + head & 157 & 0.7271 & $-0.029$ [$-0.041,-0.003$] \\
Normalized quadratic & 137 & 0.7355 & $-0.021$ [$-0.031,-0.002$] \\
\bottomrule
\end{tabular}
\end{table}

The held-out results provide consistent evidence for the circuit-induced parameterization under a fully disjoint evaluation protocol. SQUARE achieves the highest eight-task mean among all evaluated same-pipeline controls, outperforming LoRA and the wider MLP. Its improvements over the head-only and frozen-circuit variants further indicate that both the measured feature map and the learned circuit angles contribute to performance. SQUARE also exceeds the parameter-matched Givens-12 model and the normalized-quadratic predictor. The taskwise interval summaries are consistently negative for every comparator; because they are descriptive aggregates rather than calibrated confidence intervals for the macro-average, we use them to characterize the consistency of the observed ordering rather than to establish a universal ranking of quadratic models.

\subsection{Post-Entanglement Circuit Configuration Analysis}
\label{app:corrected_ablation}
We evaluate circuit configurations using a common trainable post-entanglement $RY$ block in every variant, so differences reflect the jointly learned rotation--entanglement design. QNLI, MNLI, and RTE are evaluated under the disjoint held-out protocol above with five shared seeds. Table~\ref{tab:corrected_ablation} reports the three-task mean. This analysis is diagnostic only: no circuit configuration was selected or retuned using the reported held-out test results, and all headline comparisons retain the prespecified ring configuration.

\begin{table}[h]
\centering\footnotesize
\caption{Circuit configuration analysis with a trainable post-entanglement $RY$ block in every row. Avg. test is mean held-out accuracy over QNLI, MNLI, and RTE with five shared seeds.}
\label{tab:corrected_ablation}
\begin{tabular}{llrrr}
\toprule
Pre-entangler rotations & Entangler & Angles & Total & Avg. test \\
\midrule
$RY$ & none & 8 & 153 & 0.7891 \\
$RY$ & linear CNOT & 8 & 153 & 0.7879 \\
$RY$ & ring CNOT & 8 & 153 & 0.7568 \\
$RYRZ$ & none & 12 & 157 & 0.7889 \\
$RYRZ$ & linear CNOT & 12 & 157 & 0.7875 \\
$RYRZ$ & ring CNOT (default) & 12 & 157 & \textbf{0.8462} \\
$RYRZ$ & all-to-all CNOT & 12 & 157 & 0.8155 \\
\bottomrule
\end{tabular}
\end{table}

The prespecified $RYRZ$--ring configuration gives the highest observed mean accuracy ($0.8462$), exceeding the corresponding non-entangling, linear-CNOT, and all-to-all variants in this three-task diagnostic. The contrast with the $RY$ rows is also informative: ring coupling is not uniformly beneficial, because the $RY$--ring configuration trails both $RY$--none and $RY$--linear. The result therefore supports the default circuit as a jointly specified rotation--topology design rather than attributing the gain to entanglement alone. Because the configurations were not selected or retuned using these held-out results, we treat the table as diagnostic evidence within QNLI, MNLI, and RTE; selecting a topology for a new benchmark still requires validation-only model selection followed by evaluation on an untouched test set.

\section{Experimental Setup} \label{app:setup}
\subsection{Dataset Construction and Splits}
\label{app:dataset_splits}
Table~\ref{tab:app_dataset_splits} summarizes the dataset construction and split information for the main experiments. Counts marked as controlled-boundary counts are nominal draws before the fixed near-boundary exclusion; actual retained counts can vary by seed and geometry and are the denominators used for per-run metrics.
Unless otherwise stated, all experiments use frozen BERT bottleneck representations.
For controlled classification experiments, labels are generated from predefined nonlinear interaction rules over frozen representations.

The exact generators for every geometry used in the decision-boundary experiments and controls are given in Appendix~\ref{app:boundary_generators}, and the GLUE rule in Eq.~\eqref{eq:controlled_label_rule}.
For reranking experiments, all methods score the same fixed candidate pool, so performance differences reflect the learned scoring function rather than candidate generation.
\begin{table*}[h]
\centering
\footnotesize
\caption{
Dataset construction and split information for the main experiments.
Controlled tasks use labels generated from nonlinear interaction rules over frozen BERT bottleneck features.
Reranking tasks use fixed candidate pools shared across methods.
}
\label{tab:app_dataset_splits}
\setlength{\tabcolsep}{4.5pt}
\renewcommand{\arraystretch}{1.12}
\resizebox{\textwidth}{!}{
\begin{tabular}{llrrrrl}
\toprule
\textbf{Experiment}
& \textbf{Data Source}
& \textbf{Train}
& \textbf{Validation}
& \textbf{Test / Eval}
& \textbf{Seeds}
& \textbf{Evaluation Unit} \\
\midrule

2D decision-boundary analysis (nominal)
& Controlled 2D bottleneck features
& 1,000
& 400
& --
& 3
& Binary example \\

GLUE-derived interaction classification
& GLUE inputs with controlled nonlinear labels
& 7,600
& 2,250
& --
& 3
& Sentence / sentence-pair example \\

Alpaca candidate reranking
& Alpaca instruction-following candidates
& 960
& 120
& 120
& 3
& Query--candidate pair \\

Disjoint held-out control suite
& Eight GLUE inputs with controlled nonlinear labels
& \multicolumn{3}{c}{Disjoint task-specific subsets}
& 5
& Sentence / sentence-pair example \\

\bottomrule
\end{tabular}
}
\end{table*}

\subsection{Controlled Decision-Boundary Data}
For the two-dimensional decision-boundary analysis, we construct a controlled binary classification dataset from two-dimensional bottleneck features.
Labels are generated by a nonlinear decision rule, allowing us to directly visualize whether each adapter can recover interaction-dependent boundaries.
We draw 1,000 training examples and 400 validation examples per seed and report mean results over three random seeds; these are nominal pre-filter counts. As specified with the generators below, points satisfying $|s-\tau|\leq0.05$ are removed separately from each split, so the retained evaluation denominator varies with seed and geometry. Accuracy is computed from retained correct/example counts for each run and then averaged across seeds; split details are summarized in Table~\ref{tab:app_boundary_splits}.

\label{app:boundary_data}

\begin{table}[h]
\centering
\scriptsize
\caption{
Split information for controlled nonlinear decision-boundary experiments.
The same split is used for qualitative boundary analysis and circuit configuration ablation.
}
\label{tab:app_boundary_splits}
\setlength{\tabcolsep}{5pt}
\renewcommand{\arraystretch}{1.10}
\begin{tabular}{lr}
\toprule
\textbf{Item} & \textbf{Value} \\
\midrule
Train examples & 1,000 \\
Validation examples & 400 \\
Input dimension & 2 \\
Label construction & Controlled nonlinear rule \\
Evaluation unit & Binary example \\
Metrics & Accuracy, F1, AUC \\
Seeds & 3 \\
\bottomrule
\end{tabular}
\end{table}

\subsection{Exact Generators for the Decision-Boundary Geometries}
\label{app:boundary_generators}
Each two-dimensional geometry is defined by a fixed score $s(x,y)$ on the standardized PCA-2 coordinates $(x,y)$ of the frozen bottleneck (jittered once per seed by $\mathcal{N}(0,0.015^2)$ noise); the label is $y_{\mathrm{lab}}=\mathbb{1}[s>\tau]$ with $\tau$ the median of $s$ over the training indices, and points with $|s-\tau|\le0.05$ are dropped from every split to avoid label noise at the boundary. With $r=\sqrt{x^2+y^2}$, $r_1=\sqrt{(x+0.85)^2+(y-0.25)^2}$, and $r_2=\sqrt{(x-0.85)^2+(y+0.25)^2}$:
\begin{align}
s_{\mathrm{cluster}}(x,y) &= e^{-((x+1.55)^2+(y-0.15)^2)/0.70} + e^{-((x-1.55)^2+(y+0.10)^2)/0.70} - 1.05\,e^{-(x^2+y^2)/0.55},\\
s_{\mathrm{ring}}(x,y) &= e^{-(r-1.35)^2/0.16} - 0.55\,e^{-r^2/0.75},\\
s_{\mathrm{checker}}(x,y) &= \sin(2.7x)\,\sin(2.7y),\\
s_{\mathrm{moons}}(x,y) &= e^{-(r_1-0.95)^2/0.07} - e^{-(r_2-0.95)^2/0.07} + 0.05\sin(2.0x),\\
s_{\mathrm{islands}}(x,y) &= e^{-((x+1.45)^2+(y+0.9)^2)/0.36} + e^{-((x-1.25)^2+(y-0.85)^2)/0.36} \nonumber\\
&\quad + e^{-((x+0.05)^2+(y-1.55)^2)/0.40} - 0.90\,e^{-((x-0.05)^2+y^2)/0.85}.
\end{align}
None of these is a degree-two polynomial or a single bilinear threshold. A direct map $q_\theta(x)$ would be unable to represent the radial-ring rule because it is scale invariant, but the evaluated boundary architecture first applies the learned nonlinear map $v=\tanh(Wx+b)$, which can encode radius into direction. The residual is therefore an empirical component of this broader architecture, not a mathematical necessity implied by the direct-map invariance. All constants are fixed and shared across methods and seeds. The checkerboard and radial-ring rules used in the main-text diagnostic (Table~\ref{tab:boundary_lowsup}) are the same functions.

\subsection{GLUE-derived Controlled Interaction Data}
\label{app:glue_controlled_data}
\iffalse
Table~\ref{tab:app_glue_controlled_distribution} summarizes the per-task dataset distribution used for the GLUE-derived controlled interaction classification.
\begin{table*}[h]
\centering
\scriptsize
\caption{
Task-wise dataset distribution for GLUE-derived controlled nonlinear interaction classification.
Original GLUE labels are not used; labels are regenerated from nonlinear interaction rules over frozen BERT bottleneck features.
We construct approximately balanced binary labels for each task and report macro-average accuracy across tasks.
}
\label{tab:app_glue_controlled_distribution}
\setlength{\tabcolsep}{5pt}
\renewcommand{\arraystretch}{1.10}
\begin{tabular}{lrrrrrl}
\toprule
\textbf{Task}
& \textbf{Train}
& \textbf{Validation}
& \textbf{Positive (\%)}
& \textbf{Negative (\%)}
& \textbf{Seeds}
& \textbf{Input Type} \\
\midrule
CoLA  & 1,000 & 300 & 50.0 & 50.0 & 3 & Single sentence \\
SST-2 & 1,000 & 300 & 50.0 & 50.0 & 3 & Single sentence \\
STS-B & 1,000 & 300 & 50.0 & 50.0 & 3 & Sentence pair \\
QQP   & 1,000 & 300 & 50.0 & 50.0 & 3 & Sentence pair \\
MNLI  & 1,000 & 300 & 50.0 & 50.0 & 3 & Sentence pair \\
QNLI  & 1,000 & 300 & 50.0 & 50.0 & 3 & Sentence pair \\
RTE   & 1,000 & 300 & 50.0 & 50.0 & 3 & Sentence pair \\
WNLI  & 600   & 150 & 50.0 & 50.0 & 3 & Sentence pair \\
\midrule
Total & 7,600 & 2,250 & -- & -- & 3 & -- \\
\bottomrule
\end{tabular}
\end{table*}
\fi

For GLUE-derived interaction classification, we discard the original GLUE labels and regenerate binary labels from nonlinear interaction rules over frozen BERT bottleneck features. The central held-out comparison uses disjoint task-specific train, validation, and test subsets with five shared seeds; preprocessing and the training-set median threshold are fitted on train only, model selection uses validation AUC, and test data are evaluated once.

This setting is designed to evaluate whether an adapter can recover imposed interaction-dependent rules rather than exploit original task-label correlations.
We construct task-balanced controlled subsets to prevent large GLUE tasks such as QQP or MNLI from dominating the evaluation.
Each task is approximately label-balanced, and the reported average is a macro-average across tasks.

For each task, we first extract frozen BERT bottleneck representations and then apply a fixed nonlinear interaction rule to generate labels.
The generated labels are balanced within each task, preventing adapters from exploiting class-frequency artifacts.
Because all methods use the same frozen features and controlled labels, task-wise accuracy directly measures how well each adapter recovers the imposed nonlinear interaction rule.

\paragraph{Exact label-generation rule.}
For every GLUE-derived task the controlled label is produced by one fixed rule, which we state verbatim so that the targets can be regenerated and their relation to SQUARE's feature family assessed. Let $h$ be the frozen bottleneck feature of an input (for sentence pairs, the encoder output of the concatenated pair). Features are standardized coordinate-wise on the training split and projected to $z\in\mathbb{R}^{16}$ by a PCA fitted on the training split ($d{=}16$); the validation split uses the same scaler and PCA. The controlled score is
\begin{equation}
\begin{aligned}
f(z) \;=\;& 1.20\,\sin(1.40\,z_1 z_2) \;+\; 0.90\,\cos\!\big(1.10\,(z_3-z_4)\big) \;+\; 0.80\,z_5 z_6 \\
&-\; 0.55\,z_7^{2} \;+\; 0.45\,\sin(z_8+z_9) \;-\; 0.35\,z_{10}z_{11} \\
&+\; 0.25\,\cos(z_{12}z_{13}) \;+\; 0.20\,\sin(z_{14}z_{15}-z_{16}),
\end{aligned}
\label{eq:controlled_label_rule}
\end{equation}
with one-indexed PCA coordinates, and the binary label is $y=\mathbb{1}[f(z)>\tau]$ with $\tau$ the median of $f$ on the training split; the same threshold is applied to validation, which need not be exactly balanced. The coefficients are fixed across tasks, backbones, seeds, and methods. The rule is not exactly degree two because it contains sinusoids and cosines of coordinate products, but it was deliberately designed to be interaction-rich and can therefore favor models that expose interaction features explicitly. It also contains scale- and sign-sensitive terms, whereas the pure SQUARE map is invariant to nonzero rescaling and global sign; performance is therefore distribution-conditional and is not evidence that the target belongs to SQUARE's hypothesis class. The eight task entries are eight frozen-representation distributions evaluated under one shared constructed rule, not eight independent semantic targets. The generator code and fixed coefficients are included in the anonymized reproducibility package.

\subsection{Operational Definitions for the Alpaca Reranking Protocol}
\label{app:rerank_definitions}
\textbf{Candidate generation.} For each Alpaca instruction (with its optional input field, formatted with the standard \texttt{\#\#\# Instruction / \#\#\# Input / \#\#\# Response} template), a frozen FLAN-T5 generator produces $K$ candidates by nucleus sampling with temperature $1.0$, top-$p$ $0.95$, repetition penalty $1.08$, no-repeat $3$-gram blocking, and at most $128$ new tokens; the pool is generated once and shared by every scorer. \textbf{Features.} Each (instruction, candidate) pair is encoded by frozen \texttt{bert-base-uncased}; the pooled pair representation is standardized on the training split and projected by a training-split PCA to $z\in\mathbb{R}^{d_r}$, then $\ell_2$-normalized. \textbf{Nonlinear consistency score.} $\rho_{i,j}=\sigma\!\big(\tilde\rho_{i,j}\big)$, where $\tilde\rho$ is the training-split $z$-score of
\begin{equation}
\begin{aligned}
\rho^{\mathrm{raw}}(z)=&\,1.20\sin(8z_1z_2)+1.00\cos(6(z_3-z_4))+0.90\sin(9z_5z_6)-0.80\cos(7z_7z_8)\\
&+0.70\sin(6(z_9+z_{10}))+0.60\cos(8(z_{11}-z_{12}))+0.50\sin(10z_{13}z_{14})\\
&-0.40\cos(9z_{15}z_{16})+0.30\sin(7z_{17}z_{18})+0.25\cos(6(z_{21}-z_{22})),
\end{aligned}
\end{equation}
and $\sigma$ is the logistic function; $\rho$ is therefore an author-specified interaction-dependent target on the frozen features, fixed before training and identical for all methods, and NonlinearScore (Eq.~\eqref{eq:nonlinearscore}) is its mean over the selected candidates. It measures whether a scorer can select candidates that satisfy a prescribed nonlinear consistency criterion; it is \emph{not} a measure of response quality judged externally. \textbf{Quality score.} $q_{i,j}=0.85\,\mathrm{sim}_{i,j}+0.15\,\rho_{i,j}$, with $\mathrm{sim}_{i,j}=0.25\,\mathrm{R1}+0.25\,\mathrm{R2}+0.50\,\mathrm{RL}$ the stemmed ROUGE-1/2/L $F$-measures of the candidate against the Alpaca reference output; for training only, the reference itself is added to each pool as an anchor candidate with $q$ increased by $2.0$. Score and HitRate are computed from $q$ as in Appendix~\ref{app:metrics}. \textbf{Scorer training.} Every scorer $f$ is trained on the training pools to regress $q$ with the loss $\mathcal{L}=\mathrm{MSE}(\sigma(f(z)),\,q)$, AdamW, $20$ epochs; at evaluation the candidate with the largest $f(z)$ in each held-out pool is selected. \textbf{Chance and oracle references.} The candidate-level rebuild contains $120$ evaluation pools with $K{=}8$. Uniform random Score and NonlinearScore are the averages of the within-pool means of $q$ and $\rho$. HitRate credits every candidate tied at the pool maximum, so its exact random expectation is $N^{-1}\sum_i m_i/K=0.1656$, where $m_i$ is the number of maximizers ($11$ pools contain ties), rather than simply $1/K$. Oracle-$q$ selects $\arg\max_jq_{i,j}$ and attains HitRate $1$; Oracle-$\rho$ separately upper-bounds NonlinearScore. These exact references are reported in Table~\ref{tab:alpaca_square_improved}.

\textbf{Anchor-loss caveat.} Because the anchor target is $q+2.0$ while $\sigma(f)\in(0,1)$, that target is unreachable. It acts as a shared margin-inducing training heuristic rather than ordinary bounded regression, and the reranking results are conditional on this uncalibrated choice.

\subsection{Alpaca Candidate Reranking Data}
\label{app:alpaca_reranking_data}
Table~\ref{tab:app_alpaca_splits} summarizes the split information for the Alpaca candidate reranking protocol.
\begin{table}[h]
\centering
\scriptsize
\caption{
Split information for Alpaca instruction-following candidate reranking.
All methods rerank the same fixed candidate set for each query.
}
\label{tab:app_alpaca_splits}
\setlength{\tabcolsep}{5pt}
\renewcommand{\arraystretch}{1.10}
\begin{tabular}{lr}
\toprule
\textbf{Item} & \textbf{Value} \\
\midrule
Train queries & 960 \\
Validation queries & 120 \\
Test queries & 120 \\
Candidates per query & 8 \\
Evaluation unit & Query--candidate pair \\
Selection rule & \(\hat{j}=\arg\max_j s_{i,j}\) \\
Seeds & 3 \\
\bottomrule
\end{tabular}
\end{table}
For Alpaca candidate reranking, each example consists of an instruction and a fixed set of candidate responses.
All methods score the same candidate pool and select the top-ranked candidate by
\[
\hat{j}=\arg\max_j s_{i,j}.
\]
We use 960 training queries, 120 validation queries, and 120 test queries.
Since the candidate set is fixed across methods, this experiment evaluates reranking quality rather than generation quality.

For each instruction, every method receives the same eight candidate responses.
Each method assigns a scalar score to every instruction--candidate pair and selects the candidate with the highest score.
Thus, improvements in BLEU, ROUGE, HitRate, and NonlinearScore reflect improved scoring and selection rather than improved response generation.

\subsection{Evaluation Metrics}
\label{app:metrics}
We use task-specific metrics depending on whether the experiment is formulated as binary classification, controlled nonlinear separation, or candidate reranking. Unless otherwise stated, all reported values are averaged over three random seeds.

\paragraph{Classification and controlled interaction tasks.}
For the two-dimensional decision-boundary task and GLUE-derived controlled interaction classification, each example has a binary label \(y_i\in\{0,1\}\) and a predicted probability \(\hat{p}_i\in[0,1]\). We obtain a hard prediction by thresholding at \(0.5\):
\begin{equation}
\hat{y}_i = \mathbbm{1}[\hat{p}_i \ge 0.5].
\end{equation}
Accuracy is defined as
\begin{equation}
\mathrm{Acc}
=
\frac{1}{N}\sum_{i=1}^{N}\mathbbm{1}[\hat{y}_i=y_i].
\end{equation}
Precision, recall, and F1 are computed from true positives (TP), false positives (FP), and false negatives (FN):
\begin{equation}
\mathrm{Precision}
=
\frac{\mathrm{TP}}{\mathrm{TP}+\mathrm{FP}},
\qquad
\mathrm{Recall}
=
\frac{\mathrm{TP}}{\mathrm{TP}+\mathrm{FN}},
\end{equation}
\begin{equation}
\mathrm{F1}
=
\frac{2\cdot \mathrm{Precision}\cdot \mathrm{Recall}}
{\mathrm{Precision}+\mathrm{Recall}}.
\end{equation}
We also report AUC, the area under the ROC curve. Let \(\mathcal{P}\) and \(\mathcal{N}\) denote the sets of positive and negative examples. AUC can be written as the probability that a randomly chosen positive example receives a higher score than a randomly chosen negative example:
\begin{equation}
\mathrm{AUC}
=
\frac{1}{|\mathcal{P}||\mathcal{N}|}
\sum_{i\in\mathcal{P}}
\sum_{j\in\mathcal{N}}
\left[
\mathbbm{1}[\hat{p}_i>\hat{p}_j]
+
\frac{1}{2}\mathbbm{1}[\hat{p}_i=\hat{p}_j]
\right].
\end{equation}
For circuit configuration analysis, we use AUC as the primary metric because it evaluates nonlinear separability without depending on a fixed classification threshold.

\paragraph{Candidate reranking metrics.}
For generation-oriented reranking, each instruction \(x_i\) is associated with a fixed candidate set
\(\mathcal{C}_i=\{c_{i,1},\ldots,c_{i,K}\}\). A reranker assigns a scalar score \(s_{i,j}\) to each candidate and selects
\begin{equation}
\hat{j}_i=\arg\max_{j} s_{i,j},
\qquad
\hat{c}_i=c_{i,\hat{j}_i}.
\end{equation}
All automatic metrics are computed on the selected response \(\hat{c}_i\).

BLEU measures n-gram precision against the reference response \(r_i\) with a brevity penalty:
\begin{equation}
\mathrm{BLEU}
=
\mathrm{BP}\cdot
\exp\left(
\sum_{n=1}^{N} w_n \log p_n
\right),
\end{equation}
where \(p_n\) is the modified \(n\)-gram precision, \(w_n\) is the weight for \(n\)-grams, and \(\mathrm{BP}\) is the brevity penalty.

ROUGE-\(n\) is reported as the stemmed n-gram overlap F-measure. Let $O_n$ be the clipped overlap count, $P_n=O_n/|\mathcal{G}_n(\hat c_i)|$, and $R_n=O_n/|\mathcal{G}_n(r_i)|$:
\begin{equation}
\mathrm{ROUGE}\text{-}n=\frac{2P_nR_n}{P_n+R_n},
\end{equation}
where \(\mathcal{G}_n(r_i)\) is the set of \(n\)-grams in the reference. ROUGE-L is computed from the longest common subsequence (LCS):
\begin{equation}
R_{\mathrm{LCS}}=
\frac{\mathrm{LCS}(\hat{c}_i,r_i)}{|r_i|},
\qquad
P_{\mathrm{LCS}}=
\frac{\mathrm{LCS}(\hat{c}_i,r_i)}{|\hat{c}_i|},
\end{equation}
\begin{equation}
\mathrm{ROUGE}\text{-}L
=
\frac{(1+\beta^2)R_{\mathrm{LCS}}P_{\mathrm{LCS}}}
{R_{\mathrm{LCS}}+\beta^2P_{\mathrm{LCS}}}.
\end{equation}

For reranking-specific evaluation, let \(q_{i,j}\) denote the automatic quality score assigned to candidate \(c_{i,j}\). The reported Score is the average quality score of the selected candidates:
\begin{equation}
\mathrm{Score}
=
\frac{1}{N}
\sum_{i=1}^{N}
q_{i,\hat{j}_i}.
\end{equation}
HitRate measures whether the reranker selects the highest-quality candidate in the pool:
\begin{equation}
\mathrm{HitRate}
=
\frac{1}{N}
\sum_{i=1}^{N}
\mathbbm{1}
\left[
\hat{j}_i\in\arg\max_{j} q_{i,j}
\right].
\end{equation}

NonlinearScore evaluates whether the selected response satisfies interaction-dependent quality criteria. Let
\(\rho_{i,j}\in[0,1]\) denote the nonlinear candidate-quality score. We now define both \(q_{i,j}\) and \(\rho_{i,j}\) operationally (Appendix~\ref{app:rerank_definitions}); in short, \(q_{i,j}\) is a fixed weighted combination of ROUGE agreement with the reference and \(\rho_{i,j}\), and \(\rho_{i,j}\) is a fixed, analytically specified nonlinear function of the candidate's frozen-feature coordinates, not a judgment by an external evaluator. We report
\begin{equation}
\label{eq:nonlinearscore}
\mathrm{NonlinearScore}
=
\frac{1}{N}
\sum_{i=1}^{N}
\rho_{i,\hat{j}_i}.
\end{equation}
Unlike lexical overlap metrics, NonlinearScore measures agreement with the author-specified interaction-dependent target over frozen coordinates; it is not an external judgment of response quality.

\subsection{Baseline Implementation}
\label{app:baseline_implementation}

Table~\ref{tab:app_baseline_hparams_compact} summarizes the hyperparameters used for all baselines and SQUARE.
For every method, the LM encoder and bottleneck projection are frozen, and only adapter-specific parameters and the task head are optimized.
This setting ensures that all methods operate on the same fixed representation space and that the comparison focuses on the adaptation module applied after the frozen bottleneck.
Unless otherwise stated, we use the same optimizer and scheduler family whenever applicable, so performance differences primarily reflect the adaptation mechanism rather than backbone updates, feature-extraction differences, or optimization protocol.
Hyperparameters are selected from the compact search ranges reported in Table~\ref{tab:app_baseline_hparams_compact}, and the final reported parameter counts include only trainable adapter and task-head parameters.

We choose baselines to cover three comparison axes.
First, BitFit, LoRA, AdaLoRA, and Prefix represent standard parameter-efficient classical adaptation mechanisms, ranging from bias-only updates to low-rank and prefix-based adaptation.
These methods provide strong references for whether interaction-dependent prediction can be recovered through conventional PEFT transformations over frozen representations.
Second, MLP serves as a classical nonlinear reference.
This baseline is important because SQUARE is designed to expose nonlinear interaction features; therefore, the relevant comparison is not only against linear or low-rank adapters, but also against a compact classical nonlinear head with a larger trainable parameter budget.
This allows us to test whether SQUARE provides competitive interaction modeling under fewer trainable parameters rather than merely outperforming restricted linear adapters.

Third, QAA and QPA represent quantum-enhanced PEFT alternatives.
QAA inserts a shallow quantum module at the activation level, while QPA uses a quantum component for parameter generation.
These baselines allow us to separate the effect of using a quantum-labeled component from the specific design choice made in SQUARE.
In contrast to QAA and QPA, SQUARE amplitude-encodes the frozen bottleneck itself and directly measures nonlinear relation features from an \(RYRZ\)--Ring--\(RY\) circuit.
Thus, the comparison evaluates not only whether quantum-enhanced adaptation is useful, but also where the quantum module is inserted and how its measured readout is used by the downstream head.

In addition to these groups, the comparison set also includes mechanism-matched classical interaction models over the same frozen bottleneck: explicit normalized quadratic feature expansions, full- and low-rank bilinear heads, factorization machines, classical Fourier features, random Fourier features, RBF-SVM, Nystr\"om features, kNN, graph-Laplacian classification, and an Isomap-based local manifold method. Each family is selected from prespecified configurations using validation data only, under the protocol-specific budgets described in Section~\ref{sec:experiments}; the boundary-specific diagnostic, the broad-family suite, and the compact-budget comparison use separate candidate configurations, and results are compared only within each protocol.

We also include SQUARE without the PQC as an internal ablation.
This variant keeps the same frozen bottleneck representation and lightweight task-head structure but removes the parameterized quantum feature map.
The ablation is intended to isolate whether the gains come from the measured quantum readout rather than from the bottleneck representation, the head capacity, or the training setup.
Together, the classical PEFT baselines, classical nonlinear baseline, quantum-enhanced PEFT baselines, and SQUARE ablation provide a controlled comparison for evaluating SQUARE as a compact nonlinear relation module over frozen LM representations.
\begin{table*}[t]
\centering
\scriptsize
\caption{
Hyperparameters used for baseline adapters and SQUARE.
Curly brackets indicate values considered during hyperparameter selection.
The frozen LM encoder and bottleneck projection are fixed for all methods.
}
\label{tab:app_baseline_hparams_compact}
\setlength{\tabcolsep}{7pt}
\renewcommand{\arraystretch}{1.05}
\begin{tabular}{lll}
\toprule
\textbf{Method} & \textbf{Hyperparameter} & \textbf{Values} \\
\midrule

BitFit
& Batch Size & \(\{16,32\}\) \\
& Optimizer & AdamW \\
& Scheduler & Linear Scheduler \\
& Learning Rate & \(\{1\mathrm{e}{-4},3\mathrm{e}{-4},1\mathrm{e}{-3}\}\) \\
& Trainable Parameters & Bias terms only \\

\midrule

LoRA
& Batch Size & \(\{16,32\}\) \\
& Optimizer & AdamW \\
& Scheduler & Linear Scheduler \\
& Learning Rate & \(\{1\mathrm{e}{-4},3\mathrm{e}{-4},1\mathrm{e}{-3}\}\) \\
& Rank \(r\) & \(\{4,8,16\}\) \\
& Scaling \(\alpha\) & 16 \\

\midrule

AdaLoRA
& Batch Size & \(\{16,32\}\) \\
& Optimizer & AdamW \\
& Scheduler & Linear Scheduler \\
& Learning Rate & \(\{1\mathrm{e}{-4},3\mathrm{e}{-4},1\mathrm{e}{-3}\}\) \\
& Initial / Target Rank & 8 / 4 \\

\midrule

Prefix
& Batch Size & \(\{16,32\}\) \\
& Optimizer & AdamW \\
& Scheduler & Linear Scheduler \\
& Learning Rate & \(\{1\mathrm{e}{-4},3\mathrm{e}{-4},1\mathrm{e}{-3}\}\) \\
& Prefix Length & \(\{8\}\) \\

\midrule

MLP
& Batch Size & \(\{16,32\}\) \\
& Optimizer & AdamW \\
& Scheduler & Linear Scheduler \\
& Learning Rate & \(\{1\mathrm{e}{-4},3\mathrm{e}{-4},1\mathrm{e}{-3}\}\) \\

\midrule

QAA
& Batch Size & \(\{16,32\}\) \\
& Optimizer & AdamW \\
& Scheduler & Linear Scheduler \\
& Learning Rate & \(\{1\mathrm{e}{-4},3\mathrm{e}{-4},1\mathrm{e}{-3}\}\) \\
& Circuit Depth & \(\{1\}\) \\
& Circuit & \(RY \rightarrow\) Ring CNOT \\
\midrule

QPA
& Batch Size & \(\{16,32\}\) \\
& Optimizer & AdamW \\
& Scheduler & Linear Scheduler \\
& Learning Rate & \(\{1\mathrm{e}{-4},3\mathrm{e}{-4},1\mathrm{e}{-3}\}\) \\
& Quantum Role & Parameter generation \\
& Representation Encoding & Not used \\
& Circuit & \(RY/RZ \rightarrow\) Ring CNOT \(\rightarrow RY\)\\

\midrule

SQUARE
& Primary held-out head & \(\mathrm{Lin}(16{\to}8)\)--\(\tanh\)--\(\mathrm{Lin}(8{\to}1)\), with biases \\
& Validation/backbone head & \(\mathrm{Lin}(20{\to}8)\)--\(\tanh\)--\(\mathrm{Lin}(8{\to}1)\), bias-free \\
& Batch Size & \(\{16,32\}\) \\
& Optimizer & AdamW \\
& Scheduler & Linear Scheduler \\
& Learning Rate & \(\{1\mathrm{e}{-4},3\mathrm{e}{-4},1\mathrm{e}{-3}\}\) \\
& Encoding & Amplitude encoding \\
& Circuit & \(RY/RZ \rightarrow\) Ring CNOT \(\rightarrow RY\) \\
& Readout & Probabilities (primary); probabilities + Pauli-\(Z\) (V20) \\
& Circuit Depth & 1 \\

\bottomrule
\end{tabular}
\end{table*}

\end{document}